\documentclass[useAMS,usenatbib]{mn2e}
\usepackage{epsfig,graphics}

\def\rhoDM{\mbox{$\langle\rho_{\rm DM}\rangle$}}
\def\Re{\mbox{$R_{\rm eff}$}}

\def\sigc{\mbox{$\sigma_0$}}

\def\Msun{\mbox{$M_\odot$}}
\def\Mvir{\mbox{$M_{\rm vir}$}}

\def\ML{\mbox{$M/L$}}
\def\Yst{\mbox{$\Upsilon_{\star}$}}

\def\Ydyn{\mbox{$\Upsilon_{\rm dyn}$}}

\def\mst{\mbox{$M_{\star}$}}
\def\age{\mbox{$\rm age$}}
\def\Zsun{\mbox{$Z_{\odot}$}}

\def\lsim{\mathrel{\rlap{\lower3.5pt\hbox{\hskip0.5pt$\sim$}}
    \raise0.5pt\hbox{$<$}}}                
\def\gsim{~\rlap{$>$}{\lower 1.0ex\hbox{$\sim$}}}

\def\fDM{\mbox{$f_{\rm DM}$}}

\def\mtot{\mbox{$M_{\rm tot}$}}
\def\kms{\mbox{\,km~s$^{-1}$}}
\def\esf{\mbox{\,$\epsilon_{\rm SF}$}}

\def\ta05{\mbox{$\tau^\star_{1/2}$}}
\def\t05{\mbox{$t^\star_{1/2}$}}
\def\fimf{\mbox{$f_{\rm IMF}$}}

\title[Early-type galaxy formation and dark matter]{The central dark matter content of early-type galaxies:
scaling relations and connections with star formation histories}

\author[Napolitano et al.]{\noindent
N.R.~Napolitano$^{1}$\thanks{E-mail: napolita@na.astro.it},
A.J. Romanowsky$^{2}$,
C.~Tortora$^{3,4}$
\\
~\\
$^1$INAF -- Osservatorio Astronomico di
Capodimonte, Salita Moiariello, 16, 80131 - Napoli, Italy\\
$^2$UCO/Lick Observatory, University of California, Santa Cruz,
CA 95064, USA\\
$^3$Universit$\ddot{a}$t Z$\ddot{u}$rich, Institut f$\ddot{u}$r Theoretische
Physik, Winterthurerstrasse 190, CH-8057, Z$\ddot{u}$rich, Switzerland\\
$^4$Dipartimento di Scienze Fisiche, Universit\`{a} di Napoli Federico II,
Compl. Univ. Monte S. Angelo, 80126 - Napoli, Italy\\
}

\begin{document}

\date{Accepted  Received }

\pagerange{\pageref{firstpage}--\pageref{lastpage}} \pubyear{2010}

\maketitle

\label{firstpage}

\begin{abstract}

We examine correlations between the masses, sizes, and star formation histories
for a large sample of low-redshift early-type galaxies,
using a simple suite of dynamical and stellar populations models.
We confirm an anti-correlation between size and stellar age,
and survey for trends with the central content of dark matter (DM).
An average relation between central DM density and galaxy size of
$\langle\rho_{\rm DM}\rangle \propto \Re^{-2}$ provides the first clear
indication of cuspy DM haloes in these galaxies---akin to standard $\Lambda$CDM
haloes that have undergone adiabatic contraction.
The DM density scales with galaxy mass as expected,
deviating from suggestions of a universal halo profile for dwarf and late-type galaxies.

We introduce a new fundamental constraint on galaxy formation by finding
that the central DM fraction decreases with stellar age.
This result is only partially explained by the size-age dependencies,
and the residual trend is in the opposite direction to basic DM halo expectations.
Therefore we suggest that there may be a connection between age and halo contraction,
and that galaxies forming earlier had stronger baryonic feedback which expanded their
haloes, or else lumpier baryonic accretion that avoided halo contraction.
An alternative explanation is a lighter initial mass function for older stellar populations.

\end{abstract}

\begin{keywords}
dark matter -- galaxies : evolution  -- galaxies : general --
galaxies : elliptical and lenticular.
\end{keywords}

\section{Introduction}\label{sec:intro}

The formational history of early-type galaxies (ETGs: ellipticals and lenticulars)
remains an outstanding question.
While these dynamically hot systems may be basically understood as end-products
of galaxy mergers, the details of these mergers and their
cosmological context are unclear.
High-redshift ($z$) observations have made initially surprising discoveries that
many ETGs were already present at early times with mature stellar populations,
and that these galaxies were much more compact than those in the present day.
(e.g. \citealt{2004Natur.430..181G,2005ApJ...626..680D,2006ApJ...650...18T}).

The evolution in ETG sizes is still controversial in both observation and interpretation
(e.g. \citealt{2009Natur.460..717V,2010ApJ...709.1018V,2010MNRAS.401..933M,2009arXiv0907.2392V}).  However, the most likely scenario is for a combination of
effects where individual galaxies grow in size by accretion of smaller, gas-poor
galaxies (an ``inside-out'' picture of galaxy formation),
and where younger ETGs are formed with larger sizes because of their decreased cold
gas content and the lower background densities of dark matter
(DM; e.g. \citealt{2009MNRAS.392..718S,2009ApJ...697.1290B,2009ApJ...698.1232V,2009ApJ...699L.178N,2010MNRAS.401.1099H,2009arXiv0912.0012S}).

The role of DM is considered fundamental to the formation of galaxies, and
DM halo properties have been extensively studied in cases such as gas-rich
spirals and nearby dwarfs which have suitable observational tracers
(e.g. \citealt{2007ApJ...659..149M,2009ApJ...704.1274W,2009arXiv0911.1998K}).
Studying DM in the general population of ETGs is in many ways more difficult,
with ongoing surveys of the available large-radius halo tracers attempting to
remedy our ignorance in this area
(e.g. \citealt{2009AJ....137.4956R,2009MNRAS.394.1249C,2009MNRAS.398...91P};
\citealt[hereafter T+09]{2009ApJ...691..770T}; \citealt{2010arXiv1002.3142W}).

DM can alternatively be studied with less precision but in more
extensive ETG samples by considering the well-studied central
regions---inside the ``effective radius'' (\Re) enclosing half the
stellar light, where DM is generally thought to be a minor yet
potentially detectable contributor to the mass. Here one of the
classic approaches is to analyze the ``fundamental plane'' (FP)
relating ETG sizes, luminosities, and central velocity dispersions
(\sigc). The FP shows a ``tilt'' or systematic deviation from
simple expectations based on galaxies with constant dynamical
mass-to-light ratios \ML{}, probably implying systematic
differences in the stellar populations or in DM content.

After many years of debate, there is still not a consensus on what
is driving the FP tilt (e.g.
\citealt{2004ApJ...600L..39T,2005ApJ...623L...5F}; \citealt[C+06
hereafter]{Cappellari06};
\citealt{2006MNRAS.369.1081B,2006MNRAS.370.1445D,2007ApJ...665L.105B,2008ApJ...684..248B,2008ApJ...678L..97J};
\citealt[hereafter paper I]{2009MNRAS.396.1132T};
\citealt{2009ApJ...702.1275A,2009PhDT.........9G,2010MNRAS.402L..67G,2010MNRAS.tmp..135H,2009arXiv0912.4558L}).
Some work suggests stellar populations variations or
non-homologies in the luminosity profiles as the tilt drivers.
However, it is a fairly generic expectation from the standard
cosmological framework for galaxy formation that the central DM
content of ETGs will systematically increase with luminosity---a
point we discussed in paper I and develop further in this paper.
If the FP tilt is {\it not} caused in large part by DM, there
could be problems implied for galaxy formation theory. Here one
could pursue two different philosophies: to empirically and
robustly determine the reasons for the FP tilt without recourse to
theory (e.g. making no assumptions about the underlying DM
profiles); or to adopt the theoretical framework as broadly
correct and consider the detailed implications for ETG composition
and formation.

Following the second approach,
it is now time to begin moving on from phenomenological questions about the FP tilt,
and to establish more direct connections between DM in ETGs and their
formational histories. In particular, the central DM content could prove crucial
to solving the size-evolution puzzles mentioned above, and more fundamentally
to understanding the assembly of ETGs.
To this end, we now extend the analysis of paper I, which combined models of
stellar dynamics and population synthesis to infer {\it total} and {\it stellar} masses
in a large sample in nearby galaxies, and thereby to analyze the FP tilt.
There have been previous suggestions that age and star formation
timescales are important fourth parameters in the FP
\citep{2002MNRAS.330..547T,2009MNRAS.397...75G,2009PhDT.........9G}.
We will now explore this possibility systematically, using the results from paper I
to consider additional correlations involving DM content and star formation histories (SFHs).

It should be noted at the outset that the data and the analysis techniques that
we use for deriving mass and SFH parameters may not be the most state-of-the-art.
However, our aim is to pioneer a
framework for interpreting any
data of this kind in a broad cosmological context (where in particular, emerging
high-$z$ data-sets provide only crude observational constraints on ETGs),
and we are so far
able to tentatively identify some basic and intriguing trends.

The paper is organized as follows.
In Section~\ref{sec:observ} we present our basic observational results.
We go on to analyze some implications of
the trends of DM with mass in Section~\ref{sec:dmimp} and with age in Section~\ref{sec:lcdm2}.
In Section~\ref{sec:concl} we summarize our conclusions.
Two Appendices include analyses of statistical and systematic uncertainties.

\section{Observational results}\label{sec:observ}

Here we present our basic observational results.
Section~\ref{sec:sample} provides an overview of our galaxy sample and mass inferences,
and summarizes the trends versus mass.
Section~\ref{sec:fdm_sfe} presents results related to galaxy age.

\subsection{Summary of sample and initial results}\label{sec:sample}

Our starting point is a collection of $335$ local ETGs from
\citet{PS96} that we recently re-analyzed in paper I. The sample was
selected to have measurements of \sigc\ and at least two colours,
with a subsample of $\sim220$ systems that also include the maximum rotation velocity
($V_{\rm max}$).
There are 218 elliptical galaxies and
117 lenticular/S0 systems: as shown in paper I, the
two subsamples show similar stellar and DM properties and we will
consider them jointly in this work.

We next summarize the main steps of the paper I analysis, starting
with the stellar population models. We used a set of simple
stellar population (SSP) synthetic spectra from the prescription
of \citet[BC03 hereafter]{BC03} to fit the observed galaxy
colours. We assumed a \citet{Salpeter55} or
\citet{Chabrier01,Chabrier02,Chabrier03} initial mass function
(IMF) alternatively, with initial masses $m$ in the range $0.1-100
\Msun$. For the current paper, we instead use an intermediate
\citet{Kroupa01} IMF as our default model\footnote{For old stellar
populations ($\gsim$1~Gyr), changing the IMF generally impacts the
colours at less than the $\sim$~0.01~mag level. The IMF change can
then be fairly represented as a simple overall stellar \ML\
renormalization, which is reduced by a factor of 1.6 in changing
from Salpeter to Kroupa; see e.g. Fig~4 of BC03.}.

These single-burst models were
convolved with an exponential SFR $\propto e^{-t/\tau}$ to generate more general SFHs,
where $\tau$ is a characteristic time scale. The \age,
metallicity ($Z$), and
$\tau$ were free parameters
of the model while stellar \ML, \Yst, was inferred from the best-fit
solution for each galaxy. The reliability of the modelling
technique and the intrinsic parameter scatter, as well as the
presence of spurious correlations among the stellar parameters
induced by the stellar modelling procedure, have been checked
through Monte Carlo simulations (see paper I for details,
as well as Appendix~\ref{sec:modsys} of this paper).
In addition, a recent novel, completely independent technique for estimating
\Yst\ using globular cluster systems has yielded results in
perfect agreement with ours (\citealt{2009MNRAS.397.1003F}, Fig.~6).

We derived the dynamical \ML\ (\Ydyn) within \Re\ by
means of Jeans analysis assuming spherical symmetry, isotropy
of the velocity dispersion tensor, and introducing a rotation velocity
correction in the spherically averaged RMS
velocity $v_{\rm RMS}=\sqrt{v^2 + \sigma^2}$, which is a measure
of the total kinetic energy of the system within \Re.
In the following we will use \Ydyn\
obtained with the
multi-component model in paper I (i.e. \citealt{Sersic68} profile for the stars
and singular isothermal sphere for the total mass); we have verified
that the particular mass model choice does not affect the conclusions of this paper.

\begin{figure}
\hspace{-0.2cm}
\psfig{file=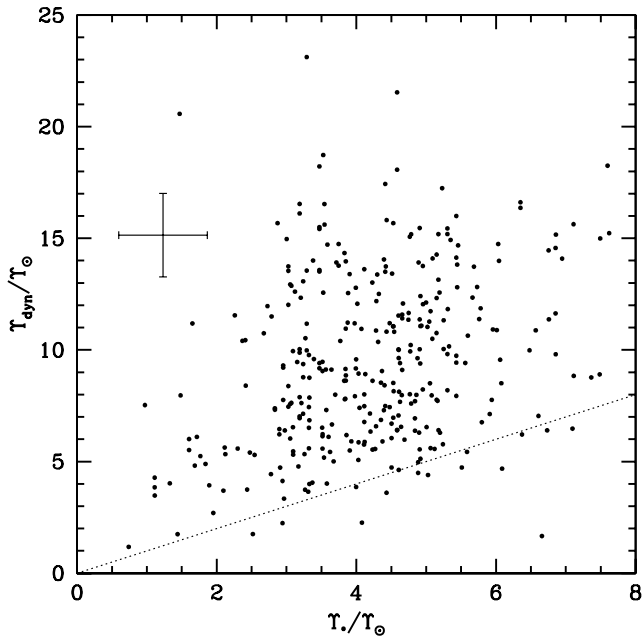, width=0.50\textwidth}
\hspace{-1.3cm} \caption{Stellar versus dynamical \ML\ within \Re\
for the early-type galaxy sample as derived in paper I,
using a Kroupa IMF.
The dotted line is the one-to-one relation,
and typical statistical uncertainties are illustrated by the error bars at left.
For most of the galaxies, the presence of a significant mass excess is evident,
implying a non-zero DM fraction in the central \Re.} \label{fig:ML}
\end{figure}

Central DM fractions are inferred by stellar and dynamical \ML\ estimates, being by
definition
\begin{equation}
\fDM=\frac{\mtot-\mst}{\mtot}=1-\frac{\mst}{\mtot}=1-\frac{\Yst}{\Ydyn},
\label{eq:fDM}
\end{equation}
where \Yst\ and \Ydyn\ are obtained in paper I\footnote{While
\Ydyn\ is a three-dimensional quantity, \Yst\ is {\it projected}
and susceptible to additional variations from large-radius
material along the line of sight. However, we have estimated using
some simple models that \Yst\ would typically be affected at only
the $\sim$~1\% level.}; the DM mass includes any mass in diffuse
gas. Note that as seen in Fig.~\ref{fig:ML}, a fraction of the
galaxies (under any of our IMF choices) appear to have $\fDM < 0$,
which would be an unphysical situation. This is not necessarily a
worry, since errors in the \ML\ estimates can scatter some data to
$\fDM < 0$. We have investigated this issue quantitatively in
Appendix~\ref{sec:unphys}, finding a strong but not definitive
suggestion that some ETGs are not compatible with having a
Salpeter IMF. Similar points were made in other studies
(\citealt{2005ASSL..327..221R}; C+06;
\citealt{2008MNRAS.383..857F,2009ApJ...704L..34C,2010arXiv1002.1083B})
using conceptually similar techniques. \citet[hereafter
T+10]{2010ApJ...709.1195T} on the other hand claimed from an
analysis of gravitational lenses that a Salpeter IMF is preferred,
with a possible systematic IMF variation. We will return to these
issues in later Sections, and for now note that an assumed
universal Kroupa IMF is a reasonable starting point.

\begin{figure}
\hspace{-0.3cm}\psfig{file=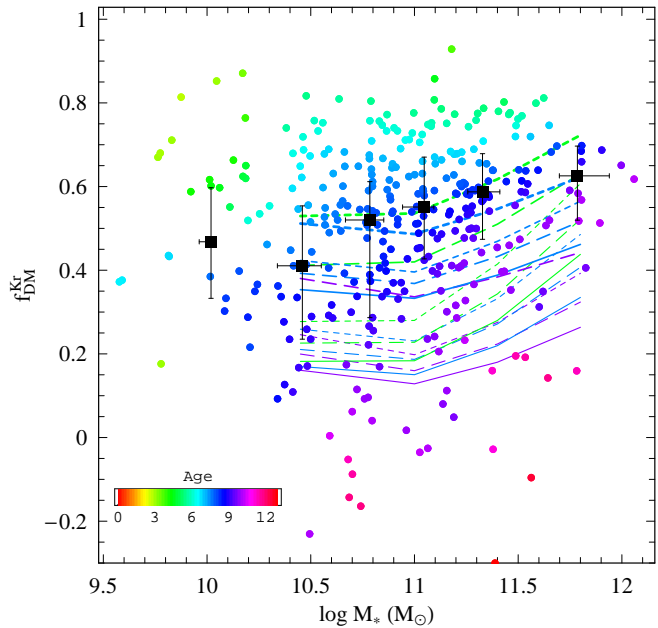, width=0.51\textwidth}
\caption{Relation between central DM fraction \fDM\ and stellar mass \mst\ for ETGs, for
an assumed Kroupa IMF.
Individual galaxy data points are shown, colourised according to their ages as illustrated by
the colour-bar at lower left; the local age variations in the plots are smoothed in
order to see the general trends.
Except for the lowest masses (which are dominated by S0s), \fDM\ increases on average with \mst
(illustrated by the black points with error bars showing the binned averages with scatter).
The curves show various predictions of $\Lambda$CDM toy models
(see Section~\ref{sec:lcdm1}).
The bottom set of 9 curves shows model predictions without adiabatic contraction (AC),
for three age bins
(3, 7, 11 Gyr, again colourised by age), and three
\esf{} values (0.7, 0.3, 0.1: solid, long-dashed, short-dashed respectively).
The top set of curves show predictions with AC, including a
smaller variation of model parameters for the sake of clarity.
Long-dashed curves show the three age bins again, with \esf=0.3 and the G+04 AC recipe.
Solid and short-dashed curves show the \esf=0.7 and 0.1 cases for G+04 AC and 7 Gyr.
The heavy long-dashed curves show \esf=0.1 for B+86 and 7 and 3 Gyr.
}
\label{fig:fDM_Mass}
\end{figure}

We found in paper I that the \fDM\ value
correlates with luminosity and stellar mass (Fig.~\ref{fig:fDM_Mass}),
a trend that could be the
{\it phenomenological} cause of the FP tilt\footnote{In
principle, the tilt could instead be caused by a systematic variation of IMF
with galaxy luminosity (see paper I).  We will reconsider IMF variations later in this paper.
The effects of ``non-homology'' of the stellar profiles were implicitly corrected for
in our modelling.}.
We also examined the
{\it formational} causes of the tilt in paper I (Section 6),
using toy models of galaxies in a cosmological context
to verify that the expected scaling relations of DM haloes might naturally
explain the \fDM\ trends.
We will develop this theme further in the following Sections, and also
begin studying the implications of the SFHs.

One basic SFH result is illustrated by Fig.~\ref{fig:downsizing}:
the stellar ages and masses of the galaxies are correlated,
providing another example of the well-known ``archaeological downsizing''
phenomenon wherein more massive systems form earliest.
We plan to consider the physical reasons for this downsizing in a subsequent paper,
comparing cosmologically-based models for SFHs and DM halo assemblies.
Here we will simply treat the SFHs as a given which we will attempt to
correlate with DM properties, envisioning these as fundamental ingredients
for constraining future theoretical models.

\subsection{New trends with age}\label{sec:fdm_sfe}

Here we move beyond the FP analysis of paper I and examine in more detail the DM
trends, including the connections with SFHs.
We plotted in Fig.~\ref{fig:fDM_Mass} the empirical values for \fDM\ versus stellar mass;
the two quantities are correlated albeit with substantial scatter and suggestions
of non-monotonicity.  We have also colour-coded the data points in Fig.~\ref{fig:fDM_Mass}
by stellar {\it age}\footnote{Hereafter, where applied, the smoothed colour coding is
obtained by linearly interpolating the average values of
the third (colour-coded) quantity over a regular grid of
the other two plotted quantities.
This procedure allows us to readily see qualitative trends in the data, with the caveat that
{\it quantifying} these trends can be unreliable after smoothing.},
and can see immediately that much of the scatter in \fDM\
might be explained by systematic correlations with SFHs.
In particular, {\it at a fixed mass the galaxies with the youngest stars are found
to have on average the highest DM fractions}.
This is a central result of our paper which we consider in more detail below.
Trends involving $\tau$ will be studied in a subsequent paper.

\begin{figure}
\hspace{-0.4cm} \psfig{file=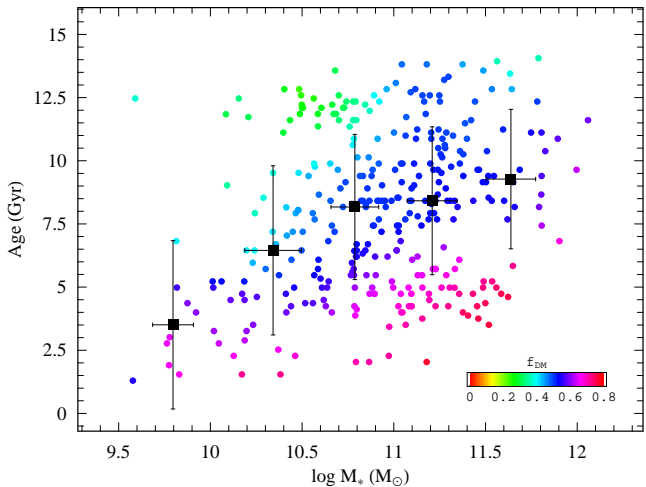,
width=0.51\textwidth} \caption{Galaxy stellar age versus
stellar mass.
The points are colour-coded according to the DM fraction (after smoothing;
see inset colour bar).
The overall average trend in mass bins is shown by points with error bars
representing the 1~$\sigma$ scatter.
}
\label{fig:downsizing}
\end{figure}

\begin{figure}
\hspace{-0.4cm} \psfig{file=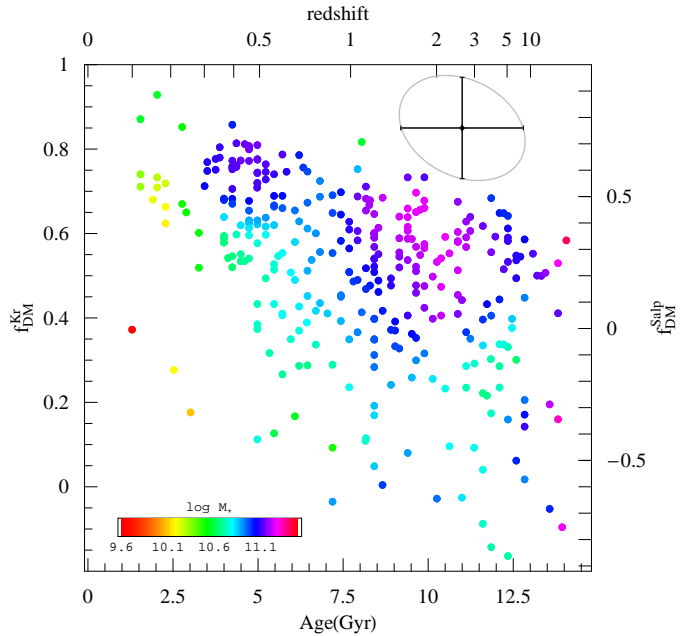, width=0.52\textwidth}
\caption{Central DM fraction of ETGs versus stellar age. The
results for a Kroupa IMF are indicated by the left-hand axis, and
for Salpeter by the right hand. The data points are colour-coded
by the locally-averaged stellar mass, with colour coding as in the
inset. A typical uncertainty ellipse for an individual galaxy (at
the centre of the range) is shown at upper right. }
\label{fig:fDM_age}
\end{figure}

We show two other projections of the three-dimensional space of age-mass-\fDM\
in Figs.~\ref{fig:downsizing} and \ref{fig:fDM_age}.
The combination of \fDM-mass and age-mass correlations (Figs.~\ref{fig:fDM_Mass} and
\ref{fig:downsizing}) suggest that there could be
an overall \fDM-age correlation.  However, Fig.~\ref{fig:fDM_age} shows that
there is a net {\it anti}-correlation between \fDM\ and age, which implies
that \fDM\ couples more strongly to age than to mass
(we will revisit this issue in Section~\ref{sec:over}).

Before analyzing these trends in more detail, we consider that there is
an important {\it fourth} parameter involved, the effective radius \Re.
This is because \fDM\ is evaluated at \Re, a variable parameter that
has its own systematic dependencies.
As we found in paper I and will revisit below,
the well-known ETG size-mass relation (\Re-\mst) can explain much
of the \fDM-\mst\ trends as an ``aperture effect'', without invoking
any intrinsic variation in the DM properties.
This is because the stellar \Re\ varies more quickly with mass than do the inner
DM halo properties (in $\Lambda$CDM), meaning that for a more massive galaxy,
the larger \Re{} encloses a higher fraction of the total DM.

The next question is whether there is a link between \Re\ and age that
could be driving the \fDM-age trends.
As mentioned in Section~\ref{sec:intro}, the size-mass relation is thought
to evolve strongly with time, such that ETGs at a fixed mass are more
compact at earlier times.
This trend can be considered generically to
include contributions both from size growth of existing galaxies,
and from the birth of new galaxies with systematically larger sizes.
As a consequence of the latter effect,
we would expect the population of ETGs at a fixed time (e.g. $z=0$) to
show a correlation between size and age at fixed mass, such that the
older galaxies are smaller.

\begin{figure}
\hspace{-0.5cm}
\epsfig{file=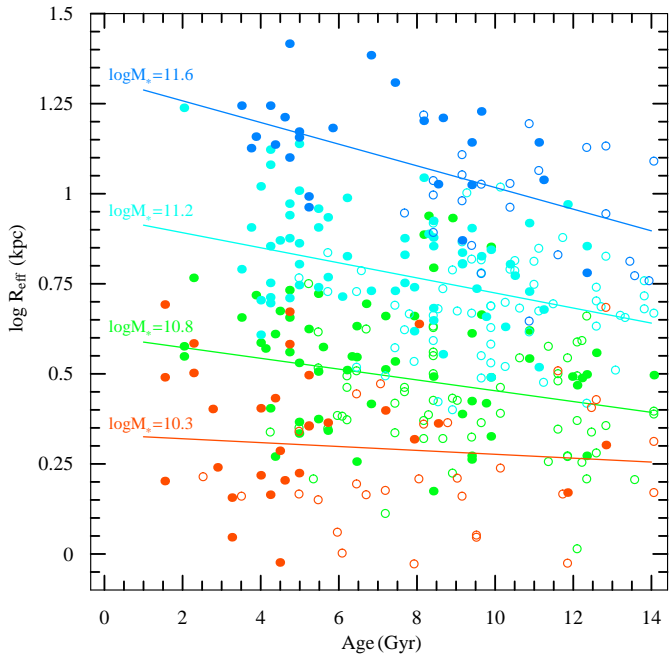, width=0.50\textwidth}
\caption{Size versus \age\ relation for ETGs at $z=0$,
with galaxies colour-coded by their stellar mass.
Solid lines show the average trends in these mass bins and are
labelled with the $\log_{10}(\mst/\Msun)$ values (Kroupa IMF).
Open and filled symbols show objects that have lower and higher \fDM{}
(respectively)
than the average for other galaxies of similar mass and age.
}
\label{fig:re_age}
\end{figure}

This qualitative prediction has recently been confirmed by
low-$z$ observations (SB09; \citealt{2009ApJ...698.1232V}),
and we now investigate it for our own ETG sample.
We plot \Re\ against age in
Fig. \ref{fig:re_age}, where
galaxies are
colour-coded according to four fiducial stellar mass bins
that we will use throughout the paper:
$\log_{10} (\mst/\Msun) \sim$~10.3, 10.8, 11.2, 11.6 for a Kroupa IMF).
The overall galaxy
distribution does not show any significant dependence of \Re\
with \age, but for any given mass
there is a clear anti-correlation
(particularly for the most massive galaxies), i.e.
older galaxies have on average smaller effective radii.
We quantify the size-age dependency by fitting a log-linear relation in each of
the mass bins.
The slopes range from $-0.005$ to $-0.030$,
where the normalization increases with mass
in accordance with the typical \Re--\mst\ relation (see e.g. paper I).
These trends appear to be independent of the
galaxy sub-type (ellipticals versus S0s)---which is a crucial
point because one might otherwise speculate that the younger objects are S0s
with systematically different properties than ellipticals.

These size-age trends might be driven by higher densities of DM and gas at
earlier epochs of galaxy formation (see references in Section~\ref{sec:intro}).
However, as with the age-mass correlation, we will for now set aside a physical
interpretation of the size-age correlation and simply consider it as a nuisance
factor that must be treated appropriately in our later toy models of DM content.
These size-age results are not meant to be universally applicable to
ETGs but are a characterization of this particular data-set as
needed for our self-consistent galaxy toy models.

The size-age trends clearly go in the right direction to explain
the \fDM-age trends:  older galaxies could exhibit lower \fDM\ simply because
their \Re\ values are smaller, so smaller volumes of their DM
haloes are probed.
In principle, one might be able to see directly from the data
whether \fDM\ is more directly coupled to age or to \Re, and to this end
we have highlighted the high- and low-\fDM\ galaxies in Fig.~\ref{fig:re_age}.
However, it is not immediately obvious whether \Re\ or age is the stronger factor,
probably because of the scatter in the data.
Instead, in the next section we will adopt a model-dependent approach,
attempting to make sense of the observed trends in the context of standard
DM+galaxy models.

Before continuing, in Appendix~\ref{sec:app} we carry out a number
of cross-checks on the robustness of our basic \fDM-age result.
First, comparing independent constraints from the literature we
find some support for an \fDM-age anti-correlation. However, there
may be disagreement in the {\it residual} \fDM-age trend after
accounting for the \Re-age relations (which we examine further in
the following Sections). Such differences illustrate the need for
further in-depth study of this topic. We also check in
Appendix~\ref{sec:app}
 the effects of our stellar populations modelling assumptions, finding
that our main conclusions are fairly insensitive to these.

Besides the systematic uncertainties, there is the issue of the statistical errors
being correlated, in the sense that an error in age correlates with an error in \Yst\
and produces an artificial \fDM-age anti-correlation (see error ellipse in Fig.~\ref{fig:fDM_age}).
To gauge this effect, we have simulated mock data-sets with random errors applied
to the ages, propagated these to errors on \Yst,
and then regenerated the \fDM-age plot.
The artificial trend generated by the errors turns out to change the slope by only $\sim -0.005$~Gyr$^{-1}$
(compared to the observed slope of $\sim -0.04$~Gyr$^{-1}$).
We will continue in the rest of this paper to assume that our empirical results
are correct, and go on to examine the implications for galaxy structure and formation.

\section{Dark matter implications}\label{sec:dmimp}

Given the preceding inferred DM trends, we begin here with interpretations of the
mass dependencies, constructing $\Lambda$CDM-based toy models
in Section~\ref{sec:lcdm1}.
We consider DM density profiles in Section~\ref{sec:profs},
and comparisons to related literature results in Section~\ref{sec:context}.

\subsection{$\Lambda$CDM halo models and trends with mass}\label{sec:lcdm1}

We now construct a series of toy models of galaxies including DM haloes based on
$\Lambda$CDM theory, as previously done in \citet{Nap05} and in paper I.
With this approach we are not modelling individual galaxies but generating
typical representatives in the space of mass and age.
Each model is spherically symmetric and is comprised of a
stellar spheroid and a DM halo.
The spheroid has a S\'ersic density distribution
with an $n$-index as described in paper I, and
an \Re-\mst-age dependence taken from the log-linear fits
in Fig. \ref{fig:re_age}.

The DM halo has an NFW density profile \citep{1997ApJ...490..493N}
following a typical concentration-virial mass ($c_{\rm vir}$-$M_{\rm vir}$)
relation\footnote{As in our previous papers, we have adopted a theoretical relation based
on a DM power spectrum normalization of $\sigma_8=0.9$ \citep{2001MNRAS.321..559B}.
We have also derived models based on a more recent estimate of $\sigma_8=0.8$
\citep{2008MNRAS.391.1940M},
which predict lower central DM content, but only at the level of changing \fDM\ by $\sim -0.05$.}.
The DM halo is optionally ``adiabatically contracted'', which is an approximate way to
model the expected drag of dissipatively infalling baryons on the surrounding DM,
producing a halo with a higher central DM density than in collisionless $N$-body simulations.

The two AC recipes we use are the most common ones, from
\citet[hereafter B+86]{Blumenthal+86} and \citet[hereafter
G+04]{Gnedin+04}, with the latter providing a weaker (and probably
more realistic) effect. Other recipes are also available (e.g.
\citealt{2005ApJ...634...70S,2006PhRvD..74l3522G}), but a more
important caveat is that the entire slow, smooth-infall AC
scenario may not be correct for forming cosmological
structures---particularly the ETGs which may have experienced
relatively violent, clumpy assembly histories. Current
high-resolution simulations do suggest that the standard AC models
are not accurate in detail, and that the baryonic effects on DM
haloes might be highly variable and stochastic
\citep{2009arXiv0902.2477A,2009MNRAS.395L..57P,2010MNRAS.402..776P,2009arXiv0911.2316T,2010arXiv1001.3447D}.
For simplicity, we will consider our AC and no-AC models as
representing two extreme models of galaxy formation, where the
baryons have alternatively strong or weak net impact on the DM
halo (cf. \citealt{2010ApJ...712...88L}).

In order to match the galaxies to their haloes, we must relate their stellar masses to their
total (or virial) halo masses.  We can parameterize this connection by assuming the cosmological
baryonic fraction, $f_{\rm bar}=\Omega_{\rm bar}/\Omega_{\rm m}=0.17$ \citep{WMAP}, is
conserved within each halo.  The stellar mass is then given by an efficiency of star formation
$\esf$, so that $\mst\ = 0.17 \esf M_{\rm vir}$.
We use a series of plausible values $\esf=(0.1,0.3,0.7)$
(e.g. \citealt{2007ApJ...667..760Z,2008ApJ...682..937B,2009MNRAS.392..801M,2009ApJ...696..620C,2010ApJ...710..903M,2009arXiv0909.4305G,2009arXiv0911.2236L};
\citealt[hereafter S+10]{2009arXiv0911.2260S}).

\begin{figure}
\hspace{-0.2cm}\psfig{file=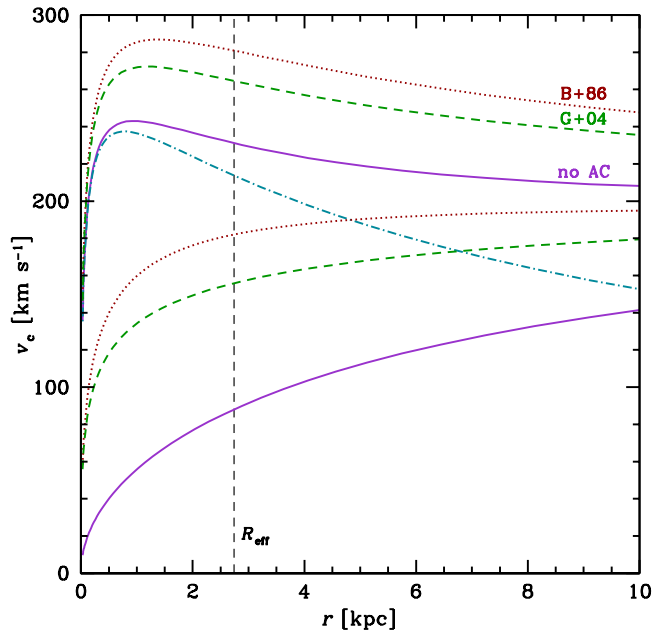, width=0.50\textwidth}
\caption{Circular velocity profiles with radius, $v_{\rm c}(r)\equiv G M(r)/r$, for cosmologically-motivated
galaxy toy models.
The stellar and halo masses are fixed to
$\log_{10}(\mst/\Msun)=10.85$ and
$\log_{10}(\Mvir/\Msun)=12.14$, respectively, corresponding to $\esf=0.3$.
Three cases are considered with different adiabatic contraction recipes, and are
indicated by curves of different colours and linestyles, labelled by each recipe.
For each case, a mass decomposition is shown, with the {\it DM} contribution as the bottom three curves,
the {\it stellar} contribution as the intermediate steeply-declining blue dot-dashed curve (the same for all cases),
and the {\it total} as the top three curves.
A vertical dashed line shows the value of the effective radius adopted.
}
\label{fig:vcprofs}
\end{figure}

We also adopt the Kroupa IMF as our fiducial choice but consider
the impact of Salpeter or Chabrier IMFs. Our model has effectively
three free ``parameters'': \esf, adiabatic contraction (AC) recipe
and IMF choice. Fig.~\ref{fig:vcprofs} shows an example of a
series of galaxy toy models for fixed stellar halo masses, and
varying AC assumptions. It can be seen that the DM content inside
1 \Re\ varies most by changing from a no-AC model to an AC model.
The choice between the B+86 and G+04 recipes makes only a small
difference on these scales\footnote{The AC recipe differences
become more important at smaller radii, where the detailed
treatment of the stellar mass profile also becomes more important,
e.g. Hernquist vis-a-vis S\'ersic models.}. This model may seem
simple, but as far as we know {\it it has never been applied to
the FP before}, except for the related study of S+10 as we discuss
later.

We now begin by deriving the predicted DM fraction within \Re\ as a function of mass,
in bins of constant age (i.e. fixing the \Re\ values for a given mass),
and with a number of different assumptions on \esf\ and AC recipe.
The results are shown in Fig.~\ref{fig:fDM_Mass}:
it can be seen first of all that without AC, the models generally underpredict \fDM.
The AC models on the other hand match the data much better {\it on average},
including reasonable reproductions of the trend with \mst.;
there is no strong preference between the B+86 and G+04 recipes.
(The data clearly show a large variation around the models, part of which is just
observational noise, and part of which is an important systematic variation that
we will examine in Section~\ref{sec:lcdm2}.)
These conclusions are IMF dependent:  with a Salpeter IMF, the \fDM\ data points all shift
downwards to lower values, and the no-AC model is preferred on average\footnote{In paper I
we focussed on the Salpeter IMF and stated erroneously that the consistency of the data with
our initial no-AC toy models was independent of IMF, when examining
the relation between mass and mean DM density within 2~\Re.
This relation for the {\it data} is indeed unaffected by IMF changes since both
the implied masses and the densities vary with the IMF.
However, we neglected to consider the effects on the {\it models}, where only the
predicted densities change (because the \Re\ values at a fixed mass are changed: see
further discussion in Section~\ref{sec:profs}).}
(although there are suspicions of Salpeter not being a valid alternative:
see Section~\ref{sec:sample}).

\begin{figure}
\hspace{-0.5cm}\psfig{file=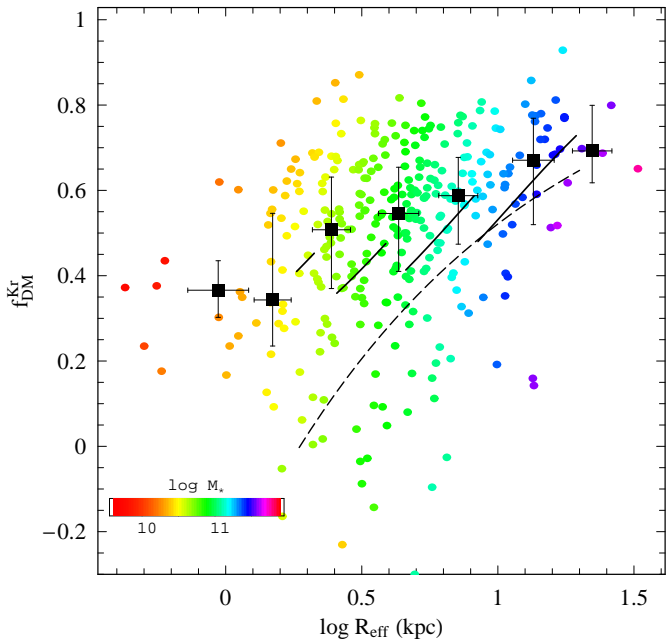, width=0.51\textwidth}
\caption{Central DM fraction versus effective radius.
Black boxes with error bars show the average trend in bins, while smaller points
show individual galaxies, colourised by the (smoothed) stellar mass according to
the inset colour bar.
\fDM\ correlates with both \mst\ and \Re\, which are themselves correlated
quantities, but the tighter trend in this Figure compared with Fig.~\ref{fig:fDM_Mass}
indicates that \Re\ is the more important parameter driving the FP tilt.
Sample $\Lambda$CDM toy models in bins of constant \mst\ (assuming $\esf=0.1$ and G+04 AC)
are shown as solid curves:  an increase of \fDM\ with \Re\ is naturally expected
in these models.
The dashed curve shows the trend found by \citet{2010MNRAS.tmp..135H}.
}
\label{fig:fDM_Re}
\end{figure}

For any fixed IMF, we see that it is a fairly generic expectation
in a $\Lambda$CDM context for \fDM\ to increase systematically
with \mst, and to produce a DM-driven tilt in the FP. The reason
is the aperture effect mentioned earlier: the scale-lengths of
galaxies' stellar parts change more rapidly than for their haloes,
causing the \Re\ region to encompass increasingly large amounts of
DM. This interpretation is supported by the \fDM-\Re\ trends in
Fig.~\ref{fig:fDM_Re} and by various FP analyses in the literature
based on simulations of galaxy formation
\citep{2005MNRAS.362..184B,2006MNRAS.369.1081B,2005ApJ...632L..57O,2006MNRAS.373..503O,2006ApJ...641...21R,2008ApJ...689...17H,2008PhDT........13C}.
A similar point from an observational basis is made by Fig.~5 of
\citet{2010MNRAS.tmp..135H}, and we include their results for
comparison in Fig.~\ref{fig:fDM_Re}; here the absolute values of
\fDM\ cannot be readily compared since their \Re\ values come from
the $K$-band and should be systematically different from ours in
the $B$-band, but their general trend with size seems comparable
to ours. The role of \Re\ in more model-dependent DM scalings is
considered further in Section~\ref{sec:profs}.

Systematic trends in \esf\ can affect the slope of \fDM-\mst\ in either direction, but
the effect is fairly weak.
Even for a factor of 7 change in assumed halo mass, \fDM\ changes only by $\sim$~0.1;
i.e. large changes in overall halo mass do not translate to large changes in central DM content.
To cancel out the DM-induced tilt, \esf\ would have to systematically increase with mass, which
goes in the opposite direction to conventional findings, wherein the most massive systems
have the highest virial $M/L$.
If the FP tilt turns out to {\it not} be driven by \fDM\ trends
(as found in \citealt{2004ApJ...600L..39T,2008ApJ...678L..97J,2009ApJ...702.1275A}),
then this would be inconsistent with vanilla $\Lambda$CDM expectations
(i.e., where central DM densities scale slowly relative to the observed stellar densities).

\subsection{Dark matter profiles}\label{sec:profs}

We next consider some interesting implications for DM density profiles with radius.
It is notoriously difficult to determine the DM content in the
centres of ETGs, and scant few studies to-date have been able to
compare the empirical DM properties to $\Lambda$CDM theory.
T+09 and paper I (Fig.~14) have found mean central densities
for DM haloes that scale with mass roughly as expected for NFW haloes.
Here we add the finding that the DM content also scales (to first
order) with galaxy size as expected.

\begin{figure}
\epsfig{file=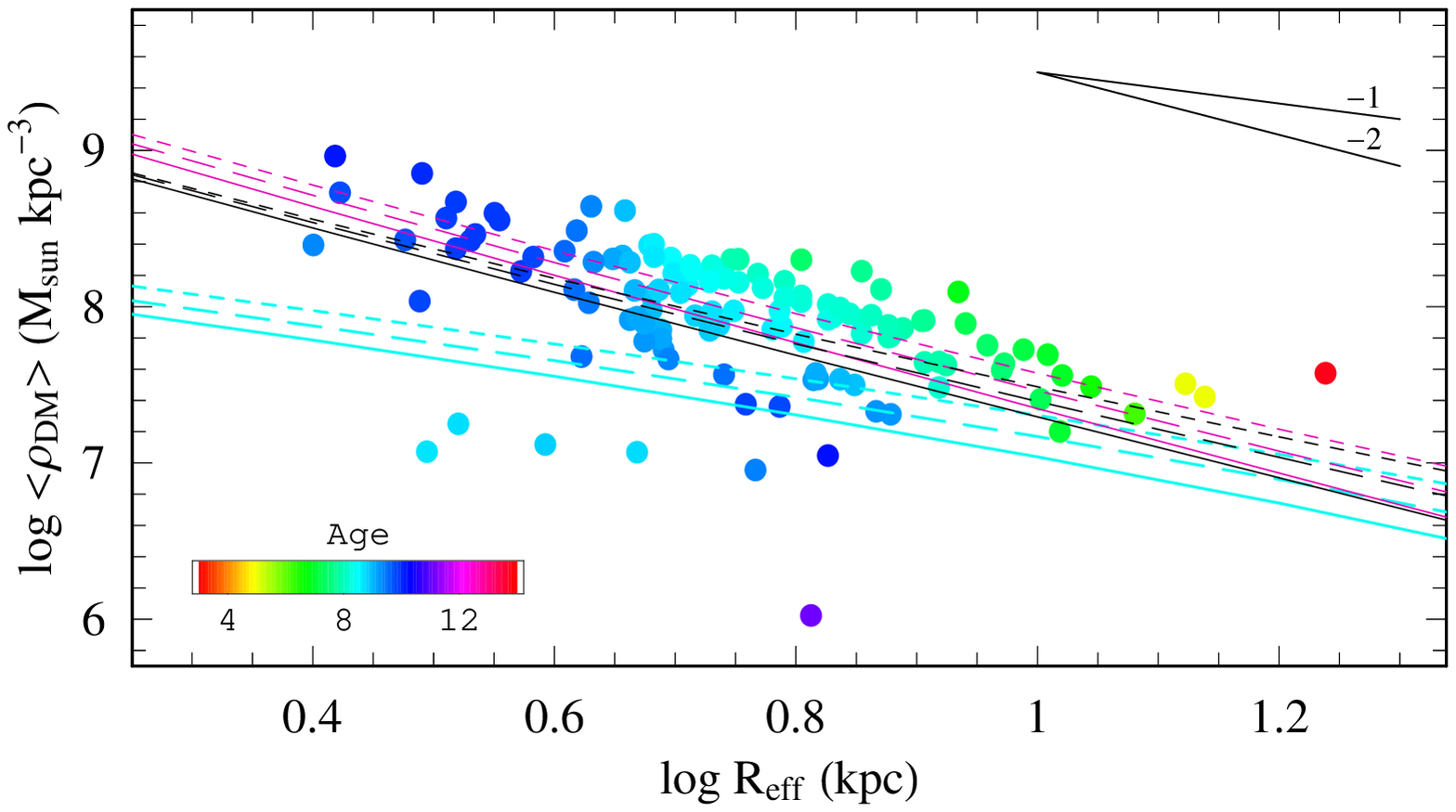, width=0.49\textwidth}
\epsfig{file=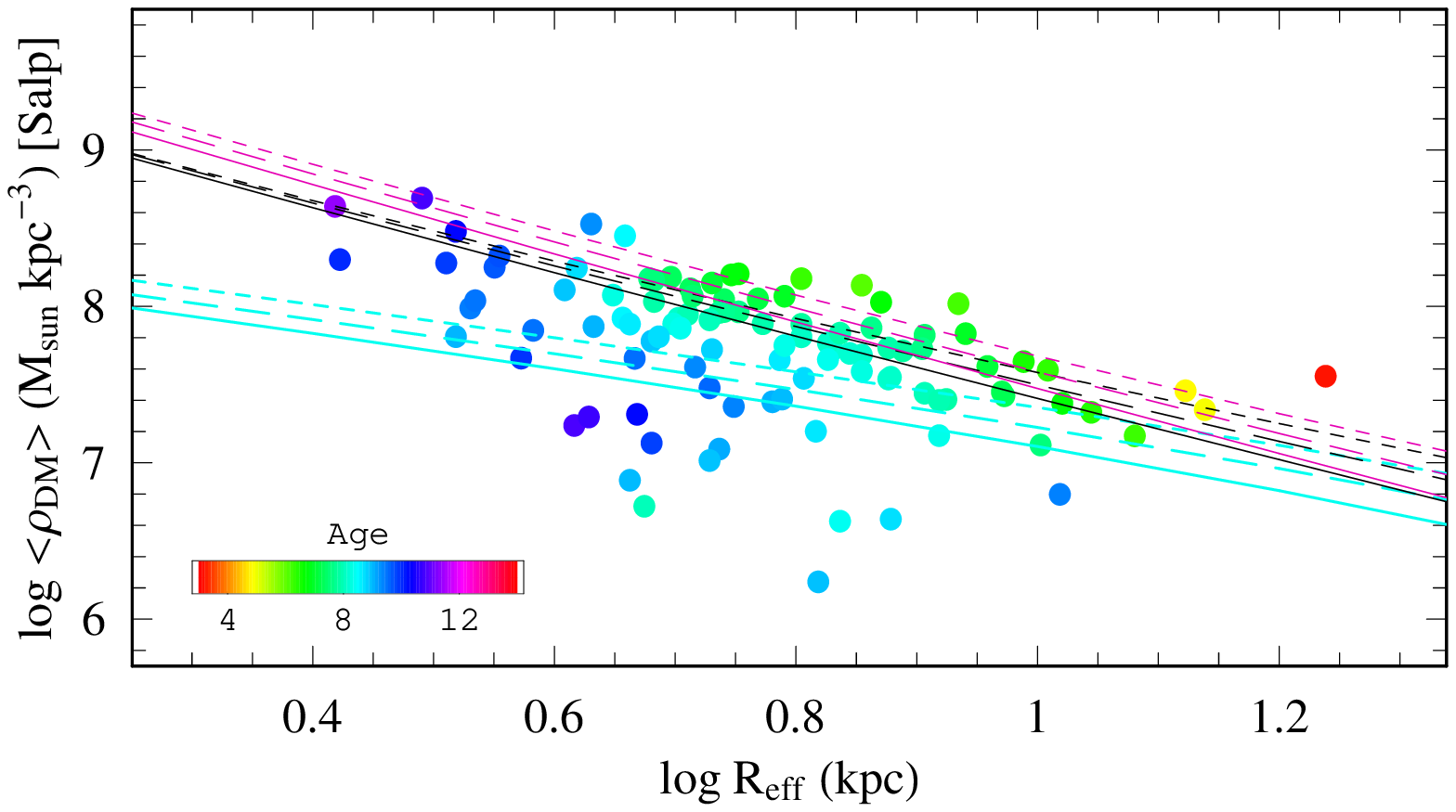, width=0.49\textwidth} \caption{Relation
between DM density within 1~\Re\, and the \Re\ value for ETGs in
one mass bin. The top panel shows the case of a Kroupa IMF, where
the mass corresponds to $\log_{10} (\mst/\Msun) \sim 11.25$, and
the bottom panel shows a Salpeter IMF for the same galaxies
[$\log_{10} (\mst/\Msun) \sim 11.4$]. Data for individual galaxies
are shown as points coloured by their ages (after smoothing: see
inset colour-bars). $\Lambda$CDM-based models are shown as curves:
the lower blue curves are for no-AC, the middle black ones for
G+04 AC, and the top red ones for B+86 AC. For each set of AC
models, fixed values of \esf$=0.7,~0.3,~0.1$ are shown as solid,
long-dashed and short-dashed curves respectively. Lines in the
upper right of the top panel illustrate power-law slopes of
$\alpha = -1$ and $-2$. The scatter in points (including the
apparent plume to very low densities) could be entirely due to
observational errors; see Fig.~\ref{fig:neg}.} \label{fig:rhoRe}
\end{figure}

We will see in Section~\ref{sec:over} that much of the observed
\fDM-age anti-correlation can be traced to the \Re-age
anti-correlation---implying that our $\Lambda$CDM toy models are
at least roughly correct. This point is illustrated more clearly
in Fig.~\ref{fig:rhoRe}, where for one example mass bin, the
stellar mass has been subtracted from the total mass, and the
residual DM content is quantified as the mean value within \Re,
\rhoDM\footnote{Note this is different from the analysis of
\citet{2006AJ....132.2711G} who looked at \rhoDM\ inside the
effective radius of the {\it DM} rather than of the {\it stars}.
Their derived slopes reflect the steeply declining outer regions
of the DM haloes, while ours involve the central regions.}.

By measuring the DM content of individual galaxies with similar masses at different radii,
we are effectively able to map out a mean DM density profile over a factor of 3 in radius.
This trick of studying composite profiles is formally valid
only if the binned galaxies all have approximately the same DM profile---as might be the
case for a no-AC NFW model.
The condition is clearly violated for models that include AC since the contraction varies among
the differently-sized galaxies.
However, the impact of this non-homology is reasonably small as discussed further below,
such that we can get an approximate idea of the DM slopes,
as well as more firmly test the null hypothesis of a universal NFW profile\footnote{One might
suspect that at some level we are getting out what we are putting in, since our default
dynamical models used in deriving \fDM\ assume an $\alpha=2$ {\it total} density profile
in order to extrapolate the \sigc\ measurements to \Re.
However, we have confirmed that using the alternative constant-$M/L$ profile yields similar
results ($\alpha$ unchanged for Salpeter IMF, and {\it steeper} by $\Delta \alpha \sim$~0.1--0.2 for Kroupa IMF).
A galaxy sample with direct mass constraints nearer to \Re\
(e.g. C+06; \citealt{2009ApJ...705.1099A,2009PhDT.........9G})
would be ideal for pursuing the density profile analysis more robustly.
We have also checked that these results are not sensitive to the stellar populations
models (Appendix~\ref{sec:modsys}).
}.

The first point to notice from Fig.~\ref{fig:rhoRe} is that the data for
\rhoDM\ and \fDM\ (see Fig.~\ref{fig:fDM_Re}) have opposite trends versus \Re,
which turns out to be crucial evidence for cuspy DM haloes on scales of $\sim$~2--15~kpc.
As \Re\ increases for a fixed galaxy mass and halo profile,
the amount of enclosed DM and therefore \fDM\ both increase.
On the other hand, the local DM density of
a cuspy halo decreases with radius, so a larger \Re\ means a smaller \rhoDM.

More quantitatively,
we may consider a general power-law density scaling for the central DM halo,
$\rho_{\rm DM}(r) \sim r^{-\alpha}$.
For $\alpha < 3$, $M_{\rm DM}(r) \sim r^{3-\alpha}$ and
therefore $\rhoDM \sim \Re^{-3} M_{\rm DM}(r=\Re) \sim \Re^{-\alpha}$.
For an uncontracted NFW halo, $\alpha\sim 1$ at the smallest radii, but near \Re\ in our toy models,
$\alpha \sim$~1.1--1.3 (steeper for higher masses because of their larger radii;
see also \citealt{2006AJ....132.2701G}).
AC effectively makes the DM even cuspier, with $\alpha \sim$~1.6--1.8 for the G+04 recipe,
and $\alpha \sim$~1.8--1.9 for B+86 (see also Fig.~\ref{fig:vcprofs}).
The effect of constructing composite profiles from non-universal haloes is for
the apparent slopes to be steeper by $\Delta \alpha \sim 0.3$.
This offset can be subtracted from the observational results to estimate the true slopes
(see below),
while more rigorous tests are made by constructing composite toy models for comparison to the data.

\begin{figure}
\hspace{-0.5cm}\epsfig{file=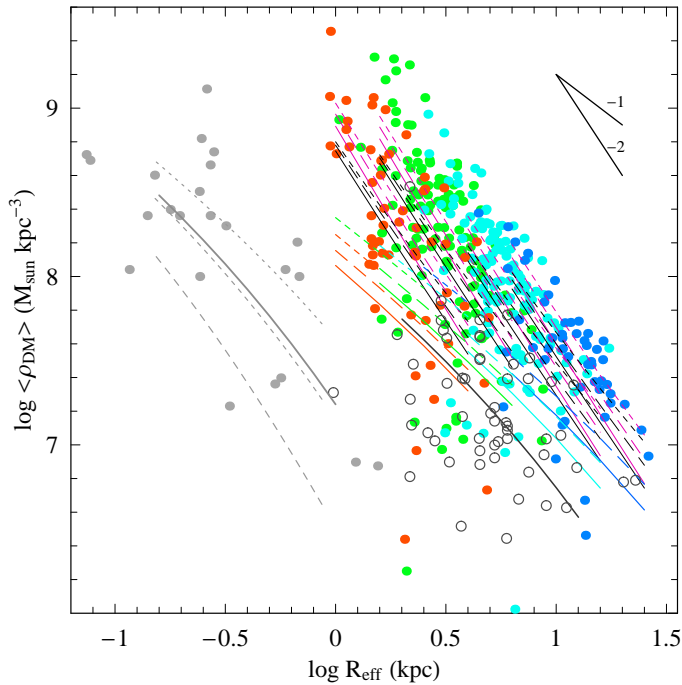, width=0.51\textwidth}
\caption{As Fig.~\ref{fig:rhoRe}, but with all ETG mass bins
plotted together (using a Kroupa IMF: see Fig~\ref{fig:re_age} for
colour key). In addition, data are shown for dwarf spheroidal
galaxies (grey filled circles on the left;
\citealt{2009ApJ...704.1274W,2010ApJ...710..886W}) and for
late-type disk galaxies (black open circles;
\citealt{2007ApJ...659..149M}). The $\Lambda$CDM toy model curves
are as in Fig.~\ref{fig:rhoRe}, with the addition of sample grey
and black model curves for the dwarf and disk galaxies. The dwarf
models include a no-AC halo sequence with $\esf=$~0.04, 0.004,
0.0004, along with a case of $\esf=0.004$ with AC (G+04 recipe;
solid curve). The spiral model uses a no-AC halo with $\esf=$~0.3.
} \label{fig:densize}
\end{figure}

Our observational results for a Kroupa IMF are illustrated by Fig.~\ref{fig:rhoRe} (top).
The vast majority of the data points in this fixed mass bin do appear to clump around
a common DM mass profile, with a composite slope of $\alpha \sim 1.9$
(so we may infer a true slope of $\alpha_0 \sim 1.6$).
A small minority of galaxies may follow a shallower trend, with a residual trend of
DM fraction versus age that is barely visible by the colourisation of the data points
(to be discussed in Section~\ref{sec:over}).
The results for the other mass bins are similar, with $\alpha \sim 1.7$--2.1,
and suggestions of a shallower slope with increasing mass (see Fig.~\ref{fig:densize}),
although it is hard to tell for sure
since the exact slopes are sensitive to the details of the fitting procedure\footnote{A
fair amount of attention in the literature has been focussed on the slopes of
{\it total} mass profiles rather than the DM slopes discussed here (e.g.
\citealt{2007ApJ...671.1568J,2009ApJ...703L..51K,2009ApJ...703.1531N,2010MNRAS.tmp..135H}).
We have not made comparisons in detail, but our toy models imply total slopes at \Re\
of $\alpha \sim$~1.8--2.2, becoming shallower with increasing mass.
This trend is driven mostly by the stellar density becoming shallower, and
is affected little by AC.}.

We also construct NFW-based model predictions as in Section~\ref{sec:lcdm1} and show them
in Fig.~\ref{fig:rhoRe}.
The steep slope for the majority of the points, along with the shallower slope with
increasing mass, turns out to correspond nicely to a range of NFW+AC models\footnote{The
overall DM density shows a tendency to increase with mass,
relative to fixed-\esf\ models (see paper I),
so the overall connection between AC and mass is not yet clear.},
with a minority of points consistent with uncontracted NFW haloes.
The model interpretations are somewhat degenerate to systematic variations of \esf\ with \Re\
(e.g. S+10 suggest an anti-correlation, which would imply the intrinsic
slopes are steeper than they appear).  However, as seen in  Fig.~\ref{fig:rhoRe},
the \esf\ effects are fairly small, and overshadowed by our fitting uncertainties anyway.

The \rhoDM-\Re\ relation is not a knock-down confirmation of $\Lambda$CDM but
intriguingly it does seem to weigh against alternative models.
If there is no DM, then of course we should have $\fDM\ = \rhoDM\ = 0$,
although there might be a systematic error in the
mass analysis, e.g. with the IMF.
In this more general case we expect $\alpha \sim 3$, and
the apparent \fDM\ to be constant with age and \Re---in
strong contradiction to the data.

The same argument may apply to, and pose a challenge for,
 alternative gravity theories that seek to explain
observational mass discrepancies at large galactic radii, usually
in spiral galaxies
(e.g., \citealt{2006ApJ...636..721B,2007MNRAS.381.1103F}).
It is beyond the scope of this paper to consider any of these theories in detail,
but we will briefly comment on the most well studied case of Modified Newtonian Dynamics
(MOND; e.g. \citealt{2007A&A...476L...1T,2008MNRAS.389..701S,2009PhRvD..80j3506F}).
In this formalism, there is a characteristic acceleration scale
$a_0 \sim 1.2\times10^{-8}$~cm~s$^{-2}$,
above which mass discrepancies should be weak or nonexistent.
In the central regions of the ETGs studied here, the accelerations are large $(\sim 4 a_0)$ and the
dynamical mass should agree well with the inferred stellar mass.

Our observed mass discrepancies, along with the striking \Re\ dependencies,
would at first glance appear to rule out MOND without DM in ETGs,
which might be related to similar results in galaxy groups and clusters
(e.g. \citealt{2003MNRAS.342..901S,2005MNRAS.364..654P,2008MNRAS.387.1470A,2008A&A...478L..23R}).
However, we cannot make a definitive statement without carrying out a
careful MONDian analysis
(see also Fig.~19 in \citealt[hereafter G+01]{2001AJ....121.1936G};
preliminary findings in \citealt{2006EAS....20..119R}; and forthcoming results in \citealt{C+10}).
In particular, the systematic uncertainties in our \Ydyn\ and \Yst\ estimates must be taken into
account, along with possible fine-tuning of the ``interpolation function'' between
the Newtonian and MONDian regimes
(cf. \citealt{2008ApJ...683..137M}).

If there are DM haloes with cored profiles (e.g. \citealt{1995ApJ...447L..25B,2010arXiv1002.0845S}),
then \fDM\ would still increase with \Re, but \rhoDM\ should be constant ($\alpha \sim 0$).
Again, using the wrong IMF could contribute an $\alpha \sim 3$ effect,
which in combination with a constant DM core could produce a shallower than
$\alpha \sim 3$ trend---but
experimenting with IMF adjustments, we cannot recover a constant-core trend.

Any of these scenarios could still be salvaged with epicycles,
e.g. with systematic errors that correlate with galaxy size,
but Ockham's razor suggests the first conclusion should be
to prefer a standard cuspy DM model.
Note that
detailed dynamical models of individual
galaxies have difficulty uniquely decomposing the stellar and DM mass profiles and thereby distinguishing
between cored and cusped DM haloes (cf. T+09; \citealt[hereafter N+09]{Nap09}).
Our approach that includes stellar populations modelling in a large galaxy sample provides the first
clear (albeit indirect) constraint on the central DM profiles of ordinary ETGs.

Before moving on, we consider the Salpeter IMF alternative,
with data and models for the sample mass bin shown in Fig.~\ref{fig:rhoRe} (bottom).
The data are seen to typically lie in between the AC and no-AC models, with a
slope of $n \sim 1.7$.
Unlike the Kroupa case where both the \fDM\ amplitude and the DM density slope
match the AC models nicely, the Salpeter case shows less overall consistency
with a simple $\Lambda$CDM model---although a caveat is that the frequent cases
with apparent negative DM densities can make the results difficult to interpret.

\subsection{Context and comparisons}\label{sec:context}

There are several recent observational analyses of DM in galaxies
that are relevant to our current study:
we will make some comparisons first to studies of ETGs, and then to other galaxy types.
We begin with two papers that are very similar in spirit to ours,
as they combine estimates of total masses and stellar populations-based masses
to derive central DM constraints in large samples of elliptical galaxies, and to
compare these to $\Lambda$CDM-based toy models.

T+10 studied a sample of strong gravitational lenses, using a combination
of lensing and dynamics to derive the total masses.
They found that a Salpeter IMF is generally consistent with a no-AC model, as did we.
They did not consider AC models explicitly, and it seems very likely that (as we have found
in this paper) lower-mass IMFs would be favoured if AC were included.

These authors also found indications of systematic changes with galaxy mass
that could be interpreted as a change either in IMF, or in the DM density slope
such that it {\it steepens} with mass ($\alpha$ increasing).
The latter effect as they mention
could be due to AC becoming stronger with mass, and appears
to go in the opposite direction of our results.
It is beyond the scope of this paper to track down the reasons for this difference
(their study did probe somewhat different regimes of radius and mass).

S+10 derived their total central masses from stellar dynamics,
and used weak gravitational lensing to derive the total halo masses and concentrations
(which we have had to {\it assume} in our models\footnote{The direct mass-concentration
results of S+10 follow well the theoretical expectations for a WMAP5 cosmology
\citep{2008MNRAS.391.1940M}, although one might expect ETGs to be
systematically biased toward higher-concentration haloes.}).
Assuming a Kroupa IMF, they found relatively high central DM masses which they showed to be
nicely consistent with G+04 AC models when breaking the sample down into
several bins in mass and in size (the latter test being implicitly equivalent
to our plots of \rhoDM\ versus \Re\ as an alternative test of the DM profiles).
Their large-radius constraints also allowed them to determine a
correlation between size and virial mass at fixed \mst, which in the context
of our formalism implies an anti-correlation between \Re\ and \esf.
If their sample has an \Re-age anti-correlation as in our sample, then
this would imply an \esf-age correlation.  We will come back to this point in
Section~\ref{sec:over}.

In addition to the qualitative consistency of the S+10 results with ours,
their quantitative DM fractions of typically $\fDM \sim$~0.6--0.7
can be compared to ours over the same mass range
[$\log_{10}(\mst/\Msun) \sim$~10.8--11.6]
where we find $\fDM \sim$~0.4--0.6.
This small difference might be accounted for by their sample selection of
only round ``central'' galaxies---which are plausibly  more DM-rich
than flattened, satellite galaxies.
These authors do note that their results seem to be somewhat high compared to
other literature studies with $\fDM \sim$~0.3--0.5
(C+06; \citealt{2007ApJ...667..176G}; see also G+01; T+09).

\citet{2007ApJ...671.1568J} analyzed a sample of galaxies using
strong lensing and stellar dynamics, and $\Lambda$CDM mass models
similar to ours. Using priors on either $\Yst$ (equivalent to a
Kroupa IMF) or total masses from weak lensing
\citep{2007ApJ...667..176G}, they find that an AC model with $\esf
\sim 0.3$ is preferred. The strong+weak lensing analysis of
\citet{2007ApJ...667..176G} on the other hand found consistency
with a no-AC model.

T+09 constructed detailed dynamical models of Coma cluster ETGs,
where \Yst\ and \Ydyn\ were determined simultaneously from the
dynamics, assuming a parameterized DM model. They found relatively
high DM densities (similar to ours), which they compared to
cosmological semi-analytic models (using no-AC models for spirals
as an additional constraint, and adopting cored DM halo models for
the ETGs which in this context tend to {\it under}-estimate the
central DM densities). They concluded that B+86 AC haloes were
preferred over no-AC.

\citet{2009ApJ...703.1257H} used X-ray emission from hot gas surrounding a small sample of
massive ETGs to constrain their mass profiles using $\Lambda$CDM-motivated models.
With a Kroupa IMF, they found that no-AC models were preferred
(although with somewhat high concentrations).
One of their sample galaxies was also the subject of a detailed dynamical study
using stars and globular clusters \citep{2009arXiv0910.4168S}, whose mass results
were systematically higher and might very well be consistent with an AC model.
This appears to be part of a significant unresolved tension between dynamical and
X-ray mass results.
On the other hand, other X-ray studies find surprisingly high DM concentrations
for galaxy groups \citep{2007ApJ...664..123B,2008MNRAS.390L..64D},
which should generally overlap with the massive end of our
ETG sample, and might support our finding that AC models are preferred.

\citet{2010ApJ...712...88L} used toy models for ``dissipationless'' and ``dissipational'' ETG formation
which should be roughly equivalent to our no-AC and AC models.
Adopting the C+06 \Yst\ values assuming a Kroupa IMF,
they found that both types of models predicted too much DM in the galaxy centres.
This disagrees with our conclusion that AC models predict about the correct amount
of DM, and no-AC models not enough.
Although there are some differences between our observational results and those of
C+06 (see Appendix~\ref{sec:modsys}),
it may also be that Lackner \& Ostriker did not allow for enough freedom in their halo
masses and concentrations to fit the data.
Our own preliminary analysis of the C+06 data suggested that a G+04 AC model would
work well on average \citep{2006EAS....20..119R}.

We next consider our ETG results in relation to other galaxy types,
showing density-size relations for different ETG mass bins,
along with data from other galaxy types in the literature,
plotted together in Fig.~\ref{fig:densize}.
We first of all consider 56 late-type galaxies from \citet{2007ApJ...659..149M},
where dynamical masses were measured by classical rotation curves,
and stellar disk masses by various methods including stellar populations modelling.
We take results from the latter (with Kroupa IMF) and then infer the DM densities within \Re,
which we define as $1.68 \times R_{\rm d}$, where the disk scale-length $R_{\rm d}$
is taken from various literature sources such as \citet{2005ApJ...632..859M}.

We plot these late-type galaxies in Fig.~\ref{fig:densize},
and see that their DM haloes are $\sim 5$ times less dense than those of early-types
at the same radii.
Qualitatively similar conclusions were reached by G+01 and T+09,
while we add the observation that
the DM density {\it slopes} are different:
the late-types have shallower slopes than the early-types ($\alpha\sim-1$ versus $\alpha\sim-2$).
The ETGs could be brought into closer consistency with the late-types if a Salpeter IMF
were adopted for the former,
but if anything a reverse IMF trend might be expected a priori (as we will discuss
further in Section~\ref{sec:IMF}).

We also show in Fig.~\ref{fig:densize} a sample $\Lambda$CDM model curve for the late-types,
for the case of $\log_{10} (\mst/\Msun) = 10$,
$\esf=0.3$ and no AC.  This curve is very similar to the equivalent one for the
lowest-mass bin of the ETGs, so for clarity we do not show additional model
curves for the late-types (whose AC model curves should differ slightly from the ETG curves).
The no-AC model appears to be preferred for the late-types\footnote{However,
\citet{2007ApJ...659..149M} among others found that $\Lambda$CDM no-AC models do not
match the data in detail, with cored DM halos being preferred.
Note that if these haloes do have constant-density cores, then there is less concern about
selecting the fairest radius for comparing their densities to ETG haloes.}, echoing the conclusions
of many other studies
\citep{2006ApJ...643..804K,2007ApJ...654...27D,2007ApJ...659..149M,2007ApJ...671.1115G,2008ApJ...684.1143X}---and contrasting with the inference of AC
for the early-types when the same IMF is adopted.

We next consider dwarf spheroidals belonging to the Milky Way, from the
compilation of \citet{2009ApJ...704.1274W} (cf. the alternative compilation of
\citealt{2009arXiv0908.2995W}).
This study extended upon earlier work that found a common DM mass within
a fixed physical radius \citep{2008Natur.454.1096S}
and suggested that these systems share a common ``universal'' DM profile with $\alpha=1.6\pm0.4$
(see also \citealt{2010arXiv1002.3376P}).
Plotting their data in Fig.~\ref{fig:densize}\footnote{The $\langle\rho\rangle$ values in
\citet{2009ApJ...704.1274W}
appear to be total mass including baryonic contributions, but this should be only a $\sim$~10\% effect.},
we note first of all that
the dwarf haloes do not appear to
join up naturally with those of more massive galaxies (cf. McGaugh et al. 2010),
and we infer that
sampling a limited range of galaxy type, mass, and radius can readily produce the impression
of a universal profile.
This conclusion is bolstered by our overplotting of a $\Lambda$CDM toy model for the dwarfs
(using a fiducial stellar mass of $\log_{10} (\mst/\Msun) = 5.5$,
Sersic index $n=0.25$ and $\esf=0.004$; a model with AC is very similar because of
the DM dominance).
For a factor of 10 variation in halo mass, the central DM densities change by
only a factor of $\sim$~3, which is less than the scatter in the data.

\begin{figure}
\hspace{-0.45cm}\epsfig{file=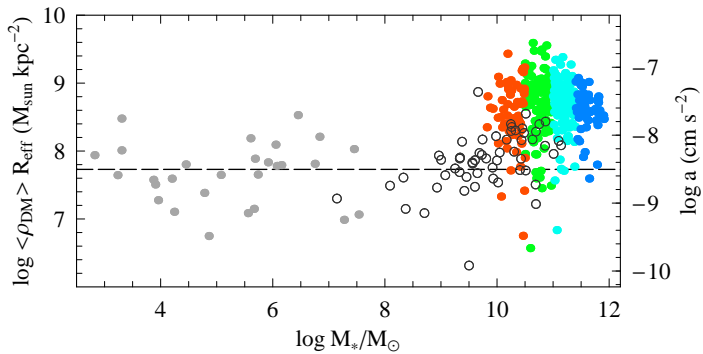, width=0.53\textwidth}
\caption{Projected density of DM in the central regions of
galaxies, versus stellar mass. The data and symbols are as in
Fig.~\ref{fig:densize}. For comparison, the projected DM density
result of \citet{2009MNRAS.397.1169D} is shown as a dashed line
(see also \citealt{2009Natur.461..627G}; our density calculation
is not exactly equivalent). The equivalent acceleration scale is
provided on the right axis; the characteristic scale in MOND is
$\log a_0/({\rm cm~s}^{-2}) \sim -7.9$. Note that these quantities
are all expressing the projection of the 3D density within \Re\
rather than a true total surface density, which would require
model-dependent extrapolations outside the regions probed by the
data. } \label{fig:surfden}
\end{figure}

Finally, motivated by the work of \citet{2009MNRAS.397.1169D} who claimed a universal DM
{\it projected} density for all galaxy types and masses (see also \citealt{2004IAUS..220..377K}),
we calculate our own approximate DM
surface density $\langle \Sigma_{\rm DM}\rangle$
from the data points in Fig.~\ref{fig:densize} by multiplying
\rhoDM\ by \Re, and showing the results versus stellar mass in Fig.~\ref{fig:surfden}.
Although our results are calculated within galaxy \Re\ rather than the DM core radius 
as in Donato et al., we reproduce their results reasonably well,
finding a fairly constant surface density on average across
a large mass range of dwarfs and late types,
but also suggesting a systematic increase with mass as
found by \citet{2009arXiv0911.3396B}.

It is beyond the scope of this paper to consider $\Lambda$CDM theoretical predictions for
the surface density in detail
(cf. \citealt{2008ApJ...686.1019Z,2009ApJ...692L.109M,2009ApJ...696.2179K,2009MNRAS.397L..87L,2009MNRAS.399L.174O,2009arXiv0911.3396B,2009arXiv0910.3211C,2009arXiv0911.1888S}).
However, the broad consistency of
the uncontracted NFW models with the dwarf and spiral data in Fig.~\ref{fig:densize} suggests that
the surface density result is not particularly surprising within the standard paradigm,
and does not
necessarily imply cored DM haloes.

We also find from Fig.~\ref{fig:surfden} that the ETG galaxies violate the constant
density scenario for the other galaxies by a factor of $\sim$~10 on average,
and a factor of $\sim$~5 in the same mass regime.
This disagrees with the conclusions of Donato et al., and we note that their
ETG results were based on a weak lensing analysis rather than central dynamical
results equivalent to those used for the other galaxies.
Other ETG dynamical results from the literature support ours
(G+01; T+09; \citealt{2009arXiv0911.3396B}; see
also the strong lensing analysis of \citealt{CT10}).

In summary, we can synthesize the constraints on the central DM content of all types of
galaxies assuming a universal Kroupa IMF, based on our study and on the literature.
We conclude that the early-type and late-type galaxies are broadly consistent with
simple $\Lambda$CDM predictions with and without AC, respectively.

The main exceptions to this emerging consensus are from X-ray studies (see above)
and from the only compilation of large-radius dynamical results (N+09),
which suggests very strong systematic AC differences among the ETGs.
However, as discussed in Appendix~\ref{sec:largeR}, selection effects are suspected
for the latter study.
We postpone further discussion of the implications of our DM results for galaxy formation to
Sections~\ref{sec:AC} and \ref{sec:IMF}.

\section{Interpreting trends with age}\label{sec:lcdm2}

At last we arrive at an analysis of the age trends.
We provide an overview of the observations compared with basic $\Lambda$CDM toy models
in Section~\ref{sec:over}.
We then attempt to explain the age trends by the systematics of DM halo concentrations
(Section~\ref{sec:con}), AC variations (Section~\ref{sec:AC}) and non-universal IMFs (Section~\ref{sec:IMF}).

\subsection{Overview of trends}\label{sec:over}

\begin{figure}
\hspace{-1.2cm}
\epsfig{file=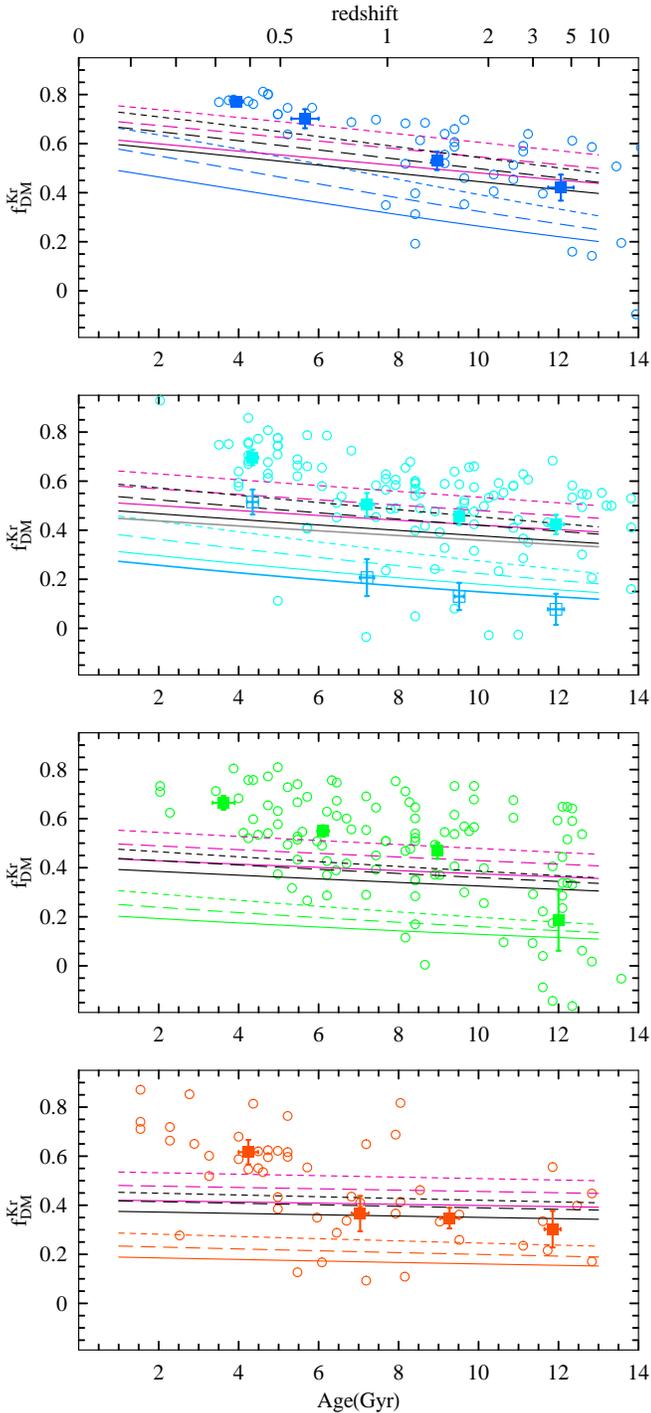, width=0.60\textwidth}
\caption{Relation between central DM fraction and stellar
age for ETGs in four mass bins:
$\log_{10} (\mst/\Msun)=11.6, 11.2, 10.8, 10.3$ from top to bottom.
Data for four age bins are plotted as points with error bars as
the mean and standard deviation for \fDM.
$\Lambda$CDM-based model predictions are shown as curves
(see Fig.~\ref{fig:rhoRe} caption for details: the model trends with age are
driven by \Re\ variations).
The default Kroupa IMF is used here, but for comparison the results for a Salpeter
IMF are included in the $\log_{10} (\mst/\Msun) =11.2$ panel:
the large open boxes show the binned data values,
and the lowest solid curve shows the $\esf=0.7$ no-AC model---which differs little
from the equivalent Kroupa curve.
Changes in the IMF affect the {\it data} much more than the {\it models}
(where the IMF indirectly links the luminosity and halo concentration).
}
\label{fig:LCDM_fdmage}
\end{figure}

As previously mentioned, the variations in \fDM\ at fixed
\mst\ appear to correlate with age, and comparison with the models in Fig.~\ref{fig:fDM_Mass}
suggests that plausible systematic variations in \esf\ with age would not be enough
to account for the trends in the data.
To see this more clearly, we construct models of \fDM\ versus age in fixed mass bins,
comparing these with the data in Fig.~\ref{fig:LCDM_fdmage}
(keeping in mind that we expect some scatter in the data points from observational
errors and from the necessary inclusion of a range of masses in each mass bin).

Both data and models agree that \fDM\ decreases with age at fixed mass,
which can be seen as a natural consequence of the \Re--\age\ trends\footnote{\citet[hereafter SB09]{2009MNRAS.396L..76S}
also found a similar trend of excess dynamical mass with age, which they
stated as driven mainly by the \Re\ variations.
However, they did not demonstrate this conclusion explicitly,
nor did they connect to specific DM models.}.
However, the data in every mass bin show even steeper age trends than predicted by
the models---particularly if we adopt models with AC, a process that couples the
DM to the baryons and  partially counteracts the trend for small stellar sizes to produce small \fDM\
(i.e. relative to uncontracted haloes, it leads to a {\it shallower} trend of \fDM\ with age).
The  difference in slope between the data ($d\fDM/d{\rm age} \sim -0.04$~Gyr$^{-1}$)
and the simple models ($\sim -0.01$~Gyr$^{-1}$ for AC,
$\sim -0.02$~Gyr$^{-1}$ for no-AC) suggests a systematic variation
with age of the DM itself.
After correcting for the effects of \Re\ (in an inevitably model-dependent way),
we support our initial conclusion from Section~\ref{sec:fdm_sfe} that
the central DM content is connected more closely to age than to mass,
since \fDM\ varies with age by up to $\sim$~0.3 relative to the model curves,
and by only $\sim$~0.1 with mass.

\begin{figure}
\hspace{-0.3cm}\epsfig{file=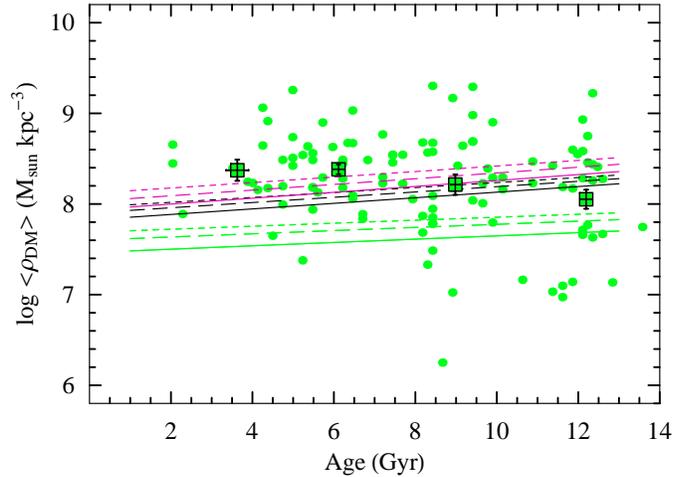,
width=0.51\textwidth}
\caption{Mean dark matter density within 1~\Re\,
as a function of stellar age, in the $\log_{10} (\mst/\Msun) =10.8$ mass bin.
Curves show model predictions as in
Fig.~\ref{fig:LCDM_fdmage}, while small points show empirical results for
individual galaxies.
Large points with error bars show the average trend in age bins.
}
\label{fig:rhoDM}
\end{figure}

We illustrate the effect of age in a different way in Fig.~\ref{fig:rhoDM} by
plotting the mean central DM density versus age, for both data and models in one mass bin.
Because of the anti-correlation of \Re\ with age, the models predict a slight
{\it increase} of \rhoDM\ with age.
However, the data show a {\it decrease} (also apparent from the colourisation in Fig.~\ref{fig:rhoRe}),
consistently with the results based on \fDM.

One obvious candidate to explain the DM-age trends would be \esf,
such that the younger galaxies simply have more massive DM haloes
overall. This scenario might be expected given cold-mode accretion
at high-$z$ \citep{2009Natur.457..451D}, where star formation is
thought to proceed very efficiently; subsequent halo growth would
have to include infall material with high \esf\, in order to not
overly dilute the net \esf\ at low-$z$. However, although it is
plausible that an \esf-age correlation contributes to the \fDM\
trends\footnote{S+10 found for a sample of low-$z$ ETGs that halo
mass correlates with \Re\ at fixed stellar mass, by a factor of
$\sim$1.5--2 in mass. If \Re\ anti-correlates with age in their
sample as in ours, then \esf\ correlates with age.}, we suspect it
is probably not the dominant factor. From examining
Figs.~\ref{fig:LCDM_fdmage} and \ref{fig:rhoDM}, systematic
variations of a factor of 10 or more in \esf\ at fixed \mst\ would
be required. While there has been a great deal of attention given
to associations of galaxies with halo masses {\it on average},
there has been much less work
on the {\it scatter} in the trends. It is outside the scope of
this paper to investigate the plausibility of order-of-magnitude
variations, but the new work of \citet{2010More}
suggests a scatter in virial $M/L$ of factors of only $\sim$~2--5
for non-satellite ETGs (increasing with \mst).

\subsection{Concentration variations}\label{sec:con}

We next consider possible connections between DM profiles and age, such that
the amount of DM within \Re\ is not directly coupled to the total halo mass.
First of all, our NFW toy models assumed the exact same mass-concentration
relation for all galaxy haloes.
However, this should really be a mean trend whose intrinsic scatter correlates
with the redshift of halo collapse $z_{\rm c}$, since halo density basically reflects the
background mass density of the universe at the time of collapse.

This concentration-age systematic can immediately be seen to go in the
wrong direction to explain the \fDM-age data,
if we make the reasonable assumption that {\it stellar} age correlates
at least weakly with the assembly age of the DM halo.
Older galaxies should therefore have {\it higher} \fDM\ and \rhoDM\
because of their denser, earlier-collapsing haloes\footnote{The stellar
component could very well also be denser in older galaxies, but we are
already accounting for this effect by including the \Re-age relations
in the toy models.}.

To illustrate this issue more quantitatively, we construct a modified
version of our $\Lambda$CDM toy models.
We begin with the simplest assumption that the age of a galaxy's stars
is directly associated with its halo's $z_{\rm c}$.
This is clearly not correct in general since in the modern picture of
``downsizing'', the star formation in massive galaxies
largely precedes the DM assembly, and vice-versa in less massive systems
(e.g. Fig.~5 of \citealt{2009ApJ...696..620C}).
However, it is beyond the scope of this paper to consider differential
assembly histories at fixed final mass, so for now we will adopt our simple prescription
as stated above.

To a good approximation, haloes of all masses and at all times
``collapse'' with a fixed concentration of $c = K \sim 4$;
the subsequent concentration evolution is basically mediated by
the expansion of the universe.
If a halo collapsed at $z=z_{\rm c}$, then at $z=0$ its concentration will
be $c \simeq K (1+z_{\rm c})$ with a scatter of $\Delta \log c \sim$~0.05--0.06
(see \citealt{2002ApJ...568...52W,2008MNRAS.391.1940M}).

\begin{figure}
\hspace{-0.3cm}
\epsfig{file=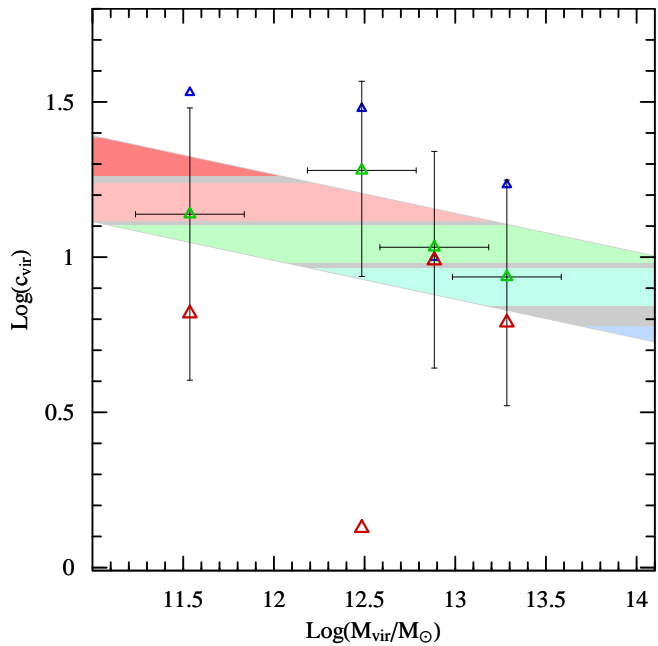, width=0.49\textwidth}
\caption{
The concentration of NFW haloes, as a function of halo mass.
The underlying grey diagonal band shows the expected average trend for 68\% of the haloes
from $\Lambda$CDM simulations.
The horizontal color bands show halo ``ages'' corresponding to collapse redshifts:
12, 11, 10, 8, 5 Gyr, or $z_{\rm c}\sim$~4, 3, 2, 1, 0.5,
for red, pink, green, cyan, blue bands (top to bottom).
The triangles with error bars show the data (assuming Kroupa IMF)
in four mass bins, with age sub-bins of 12, 8, 4 Gyr
(red, green, blue colours, respectively).
Sample error bars are included for the intermediate-age bins, showing the
scatter in masses and the 1~$\sigma$ uncertainties in the best-fit concentrations.
These uncertainties mean that the one point with a very low concentration should not
be taken too literally.
}
\label{fig:cM}
\end{figure}

This line of argument implies that {\it all galaxies with the same
stellar age have the same halo concentrations}, independent of
mass. Certainly this is an oversimplified picture and might be
undermined by late, minor gas-rich mergers biassing the inferred
stellar ages. For now we will adopt this model as an ansatz and
see where it leads us, with predictions illustrated by the model
concentration-age bands in Fig.~\ref{fig:cM}.

To derive estimates from the {\it data} for concentrations,
we proceed as follows.
First, we collect the individual galaxy results into bins of age and stellar mass,
calculating the average \fDM\ values (similar to Fig.~\ref{fig:LCDM_fdmage}).
We then fit our $\Lambda$CDM toy models (Section~\ref{sec:lcdm1}) to the \fDM\ data.
Given a fixed IMF and DM profile (NFW with or without AC), the remaining free
parameters are now \esf\ and $c$.
We fix \esf\ to be constant in each mass bin, and set it to a value that forces
the average $c$ (including all ages) to approximately agree with the theoretically
predicted mean value.
Note that for models with AC,
the $c$ value denotes the ``original'' concentration after correcting the observations for AC.

The results of this fitting exercise for a Kroupa IMF and G+04 AC prescription
are overplotted as data points in Fig.~\ref{fig:cM}.
To reproduce the mean $c$-\Mvir\ relation, we find $\esf=0.25$ for the lowest mass
bin, and $\esf=0.1$ for the others; it is noteworthy that such a simple,
plausible $\Lambda$CDM-motivated model
can be constructed to agree with both central and global DM constraints for ETGs.

Now considering the age dependencies, we see that there is a clear trend for
older galaxies to have less concentrated haloes---an
effect that is diametrically opposed to
our assumed trend for star formation to mimic halo collapse.
If taken at face value, these concentration results would imply that ETGs on average
formed their stars not long after their haloes collapsed ($z_*\sim 1$ versus $z_{\rm c}\sim 2$),
but the late-collapsing haloes tended to form their stars very early
($z_*\sim 4$ versus $z_{\rm c} \sim 0.4$)
while the early-collapsing haloes formed their stars very late
($z_*\sim 0.4$ versus $z_{\rm c} \sim 5$)---although these latter trends are more apparent
for the low-mass galaxies.
We are not aware of any physical mechanism that would produce such counterintuitive variations
{\it at a fixed mass},
and we suspect that concentration effects do not explain the observed DM-age trends.
These conclusions
are similar when adopting a Salpeter IMF with no AC.

\subsection{Adiabatic contraction}\label{sec:AC}

The next alternative for systematic differences in the DM profiles is that
the radial profiles have been systematically affected by baryons in different ways.
The simplest picture in this context is for AC strength to vary with age.
We do not explore this effect in detail, but refer to Fig.~\ref{fig:LCDM_fdmage}
for a qualitative understanding.  If the IMF has a slightly higher mass normalization
than Kroupa (or if the \esf\ values are somewhat lower than the range adopted so far),
then the youngest galaxies are consistent with the B+86 AC recipe,
with AC strength decreasing with age until the oldest galaxies have had no AC.
Variations in \esf\ would then play a minor role in the central DM trends.
In principle, we could confirm this interpretation by examining the \rhoDM\ radial
profiles for different age bins, but the data are currently too noisy for this purpose.

Is there any theoretical reason to expect AC to correlate with stellar age in this way?
Or more generally, since we do not directly confirm the AC models, why should the DM be
preferentially shifted away from the centres of ``older'' galaxies, and towards the centres
of ``younger'' galaxies, by baryonic effects?
One possibility concerns the smoothness of baryonic infall:  the standard AC model
is motivated by smooth gaseous dissipation, while clumpier collapse---both in the
baryons and in the DM---is expected to diminish or even reverse the contraction
because of ``feedback'' from dynamical friction
(e.g. \citealt{2008ApJ...681.1076D,2008ApJ...685L.105R,2009ApJ...691.1300J,2009ApJ...697L..38J}).

Alternatively, more direct baryonic feedback could occur via outflows from
AGNs and supernovae, potentially puffing up the DM halo
(e.g. \citealt{2001MNRAS.321..471B,2004MNRAS.353..829M,Mash+08,2008A&A...479..123P,2010Natur.463..203G,2010arXiv1001.3447D}).
The notion of feedback effects on DM haloes is becoming commonplace, but
the puzzle here is how to explain a systematic trend with time.
Realistic simulations of galaxy formation including baryonic processes
are only now becoming powerful enough to investigate such questions in detail.

One specific scenario worth special mention is
the emerging paradigm of high-$z$ massive galaxy formation by cold streams of DM and baryons
\citep{2009Natur.457..451D,2009ApJ...703..785D,2009MNRAS.397L..64A,2009MNRAS.395..160K,2009MNRAS.396.2332K}.
The clumpiness of the infall in this picture could effectively bypass the contraction
process (as proposed by \citealt{2008ApJ...688...67E}).
These ``wild disk'' galaxies are though to
initially form hot stellar disks and bulges in situ, and then evolve into low-$z$
ellipticals, S0s, and early-type spirals---whether by passive fading or by eventual merging
\citep{2008ApJ...679.1192C,2008ApJ...688..789G}.

We would like to connect the DM-age trends for ETGs with the
impacts of physical processes in the cosmological context of
galaxy formation. To begin doing so, let us first consider even
broader questions relating the different {\it types} of massive
galaxies. As discussed in Section~\ref{sec:context}, our initial
assessment of low-$z$ galaxies is that the early-types have haloes
with AC, and the late-types without AC. This type difference is
not simply an age difference since among the ETGs, younger
galaxies appear to have {\it stronger} AC. These observations
raise an interesting cladistical question: if ETGs are generally
formed in the mergers of spirals per the conventional wisdom, do
they have roughly the same central DM densities? Also, if the most
massive ETGs are assembled by dry mergers of smaller ETGs, are
their respective DM densities consistent with this picture?

A related problem was raised long ago for the {\it stellar}
densities of galaxies. It is well known that in the merger of
collisionless systems, the phase space density $f$ cannot be
increased, but observationally the central stellar $f$ values of
ETGs are much higher than those of spirals
\citep{1986ApJ...310..593C,1993ApJ...416..415H,1998MNRAS.296..847M}.
This puzzle was solved by taking into account the dissipational
infall of gas during a merger, forming new stars in a high-density
central starburst (e.g.
\citealt{1992ApJ...390L..53K,1998ApJ...497..108B,2000MNRAS.312..859S}).

Returning to the DM issues,
the 6D density $f$ is of course not the same as the 3D density $\rho$, or 2D $\Sigma$, but
we will not carry out a rigorous analysis of $f$ here\footnote{We have attempted
rough estimates of the coarse-grained phase-space density via
$\rhoDM \sigma_{\rm DM}^{-3}$ (e.g. \citealt{2001ApJ...561...35D}),
but the results are very sensitive to
how the DM dispersion $\sigma_{\rm DM}$ is handled, and will require
more careful analysis.}.
Instead, noting that
discrepancies involving $f$ tend to be associated with $\rho$ and $\Sigma$ problems,
we will take a very simplified approach of conserving mass:
the amount of central DM in a merger remnant should be roughly equal to the
sum of the progenitors' central DM
(e.g. \citealt{2004MNRAS.349.1117B,2006ApJ...641..647K}).

The observed DM density differences between early- and late-types appear from
Figs.~\ref{fig:densize} and \ref{fig:surfden} to be factors of $\sim$~5.
However, in this context the figures are difficult to interpret because
the two galaxy types occupy an almost disjoint region of \Re-\mst\ space.
Comparison of the small samples of overlapping objects in this space
indicate the real density differences may only be $\sim$~2 on average.

\begin{figure}
\hspace{-0.1cm}\epsfig{file=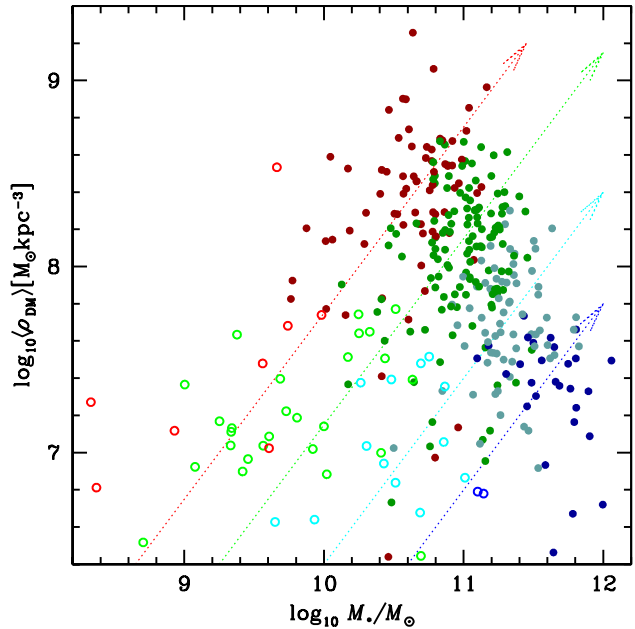, width=0.50\textwidth}
\caption{Mean DM density within \Re, versus galaxy stellar mass.
Filled dark circles are ETGs, while open bright circles are
spirals, with colours indicating bins of constant size ($\Re
\sim$~2, 4, 9, 15 kpc, from top to bottom). Arrows indicate linear
density growth with mass within the same radius. }
\label{fig:rhoM}
\end{figure}

We illustrate the trends in a different way in Fig.~\ref{fig:rhoM},
where the mean densities are plotted against stellar masses, in bins
of similar physical radius.
The simplest expectation here for galaxy mergers is that the DM density within the
same radius grows linearly with mass,
which the Figure shows is borne out remarkably well if considering the
spirals as the progenitors of the ETGs.
The full story is more complicated, as the true ETG progenitors are thought to have
been high-$z$ gas-rich systems with dense DM haloes (e.g. discussions in
Section~\ref{sec:con} and in G+01 and T+09).
Also, there are potential effects in the mergers that could raise or lower the
densities relative to the linear relation.
A detailed theoretical exploration of DM density changes in mergers would be
a useful contribution at this point, using also the constraint that the
ETG haloes appear cuspier than those of the spirals.
However, we can conclude that so far there does not appear to be an obvious
problem in generating ETG halo densities from late-type progenitors.

Now returning to the DM trends with {\it age} in the ETGs,
progenitor bias seems an unlikely explanation.
Higher DM densities in the early progenitors would produce the {\it opposite}
low-$z$ correlations to what we observe, and furthermore current observations
of candidate high-$z$ progenitors indicate relatively
{\it low} central DM content, similar to low-$z$ spirals \citep{2009arXiv0907.4777B}.

Instead, the possibility then remains
that halo contraction or expansion is a strong function of $z$.
The scenario would be for halo contraction to become more important for
ETGs that formed at later times, perhaps
in a transition from the cold stream phenomenon discussed above,
to a more merger-dominated evolution that produces stronger contraction.
Early, strong feedback might also play a role.

An additional effect to consider is dissipationless merging of ETGs: with
repeated dry mergers, the initially segregated stellar and DM profiles should
eventually converge to a well-mixed NFW-type profile for the {\it total} mass,
implying a net migration of DM toward the galaxy's centre.
One might surmise on this basis that older galaxies would have increased central DM content
because of their longer post-star-formation period of merging.
The simulations of \citet{2009ApJ...696.1094R} highlighted the effect of
\fDM\ augmentation through dry merging.
However, the main \fDM\ driver here may be the aperture effect due to the
merger-induced \Re\ growth
(as the results of \citealt{2005MNRAS.362..184B,2006MNRAS.369.1081B} suggest),
rather than a significant evolution of the DM profile itself.

\subsection{Initial mass functions}\label{sec:IMF}

After exhausting a litany of possibilities for DM properties to vary with age,
we consider the alternative that our DM inferences are wrong because we have
assumed a universal IMF when estimating the stellar $M/L$ values.
The implication would then be that the IMF varies systematically with stellar age,
in the sense that younger galaxies have a higher-mass IMF:
i.e., they have more stellar mass for a given luminosity, so the high
{\it dynamical} $M/L$ observations are due not to excess DM but to higher $M_\star$.
In this context, when we couch the IMF in terms of ``Salpeter'' and ``Kroupa'',
we will not literally mean the same detailed IMF shape as those functions, but
rather an equivalent overall \Yst\ normalization.

The subject of IMF variations is highly controversial and unresolved:
for a review see \citet{2010arXiv1001.2965B}.
There are theoretical reasons to expect lower-mass IMFs for stellar populations
that formed in the early universe
\citep{2005MNRAS.359..211L,2007MNRAS.374L..29K}, and observational suggestions for IMFs to be
lower-mass at higher $z$ (e.g. \citealt{2008ApJ...674...29V,2008MNRAS.385..147D,Holden10}).
These ideas would fit in well with the IMF-age trend we suggest above.
All these results might further be consistent with
suggestions that the IMF becomes more top-heavy
at higher SFRs \citep{2006MNRAS.365.1333W,2009MNRAS.400.1347C,2010arXiv1001.2009H},
since SFRs should have generally been higher at earlier times.
Similarly, the more compact nature of the older ETGs might imply higher gas
densities in the star-forming epoch, which has also been suggested to
produce more top-heavy IMFs
\citep{2004MNRAS.353..113L,2009ApJ...695..765M,2010arXiv1001.0971K}.

\begin{figure}
\hspace{-0.2cm}\epsfig{file=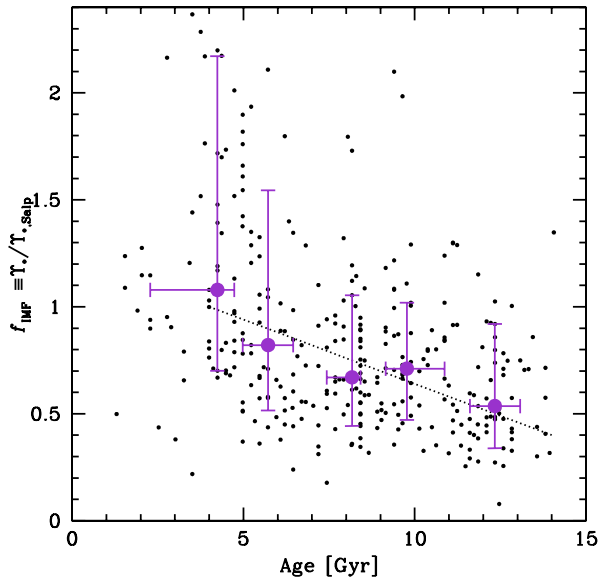, width=0.46\textwidth}
\caption{Stellar mass-to-light ratio produced by variable IMF
(relative to Salpeter), versus stellar age. The small points show
the results calculated for individual galaxies, the large points
with error bars show median values in bins, and the dotted line
shows the overall trend adopted. } \label{fig:IMF0}
\end{figure}

To be more quantitative about the implied IMF variations, we
construct a simple ad hoc model as follows. First we derive an
average \fDM-\mst\ trend based on the data
(Fig.~\ref{fig:fDM_Mass}), and for each individual galaxy, solve
for the \Yst\ value that would agree with this trend (given its
individual \Ydyn\ value). This value can be expressed relative to
the observed Salpeter-based value by a fraction $\fimf \equiv
\Yst/\Upsilon_{\star,{\rm Salp}}$ (cf. T+10). The result of this
exercise is shown in Fig.~\ref{fig:IMF0}: we see a clear trend for
the data to require a decreasing IMF mass with age, which we
roughly parametrize by a linear function of \fimf\ versus age.

\begin{figure}
\hspace{-0.4cm}\epsfig{file=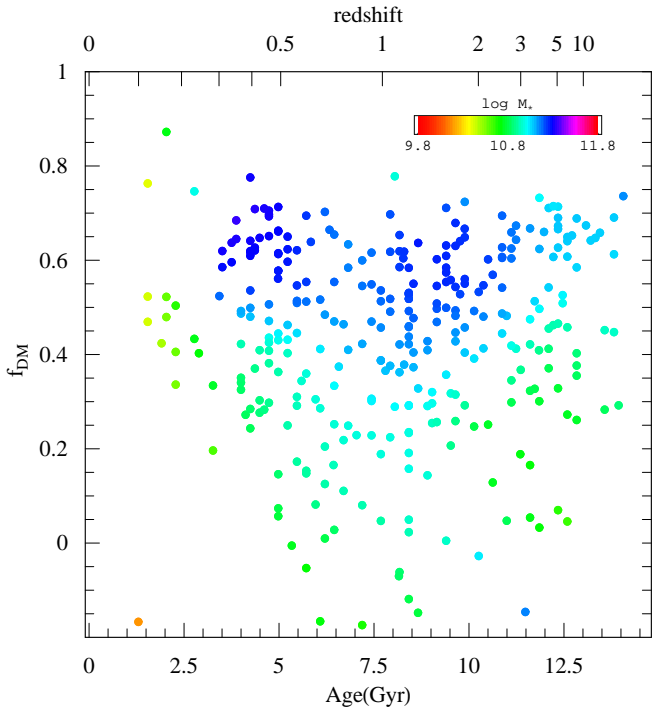, width=0.51\textwidth}
\caption{
As Fig.~\ref{fig:fDM_age}, with age-dependent IMF trend taken from Fig.~\ref{fig:IMF0}.
}
\label{fig:IMF}
\end{figure}

Given this new IMF assumption, we then re-derive some of our DM inferences.
First we show the revised \fDM-age trends in Fig.~\ref{fig:IMF}.
As intended by construction, the data show roughly constant DM fractions with age
in fixed mass bins
(and the galaxies with Salpeter IMFs are not in danger of being unphysical
since these were the ones with apparently large \fDM\ to begin with).
As we have seen earlier, the \fDM-age trends cannot be immediately
interpreted without folding in the \Re-age dependencies.
Upon investigation, we find that with the varying IMF assumption, the
strong \Re-age anti-correlations (Fig.~\ref{fig:re_age}) have disappeared,
so that there is on average no correlation.

We find it a remarkable coincidence that the same IMF model removes the
age trends in both \fDM\ and \Re.
This finding suggests that a systematic IMF variation could be real, and
that it might account in part for the apparent trends with age and redshift
of galaxy sizes (as suggested by \citealt{2009ApJ...706L.188M}).
Note that with our (somewhat arbitrarily chosen) variable IMF, the \fDM\ observations
are now generally consistent with AC halo models.

\begin{figure}
\hspace{-0.4cm}\epsfig{file=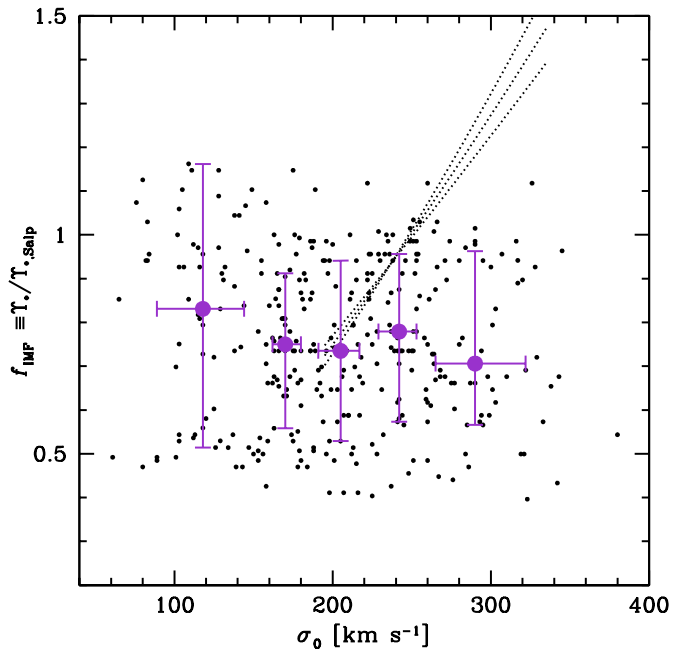, width=0.51\textwidth}
\caption{Stellar mass-to-light ratio produced by variable IMF
(relative to Salpeter), versus central velocity dispersion. The
large points with error bars show median values in bins, and the
dotted curves shows the trend found by T+10, including their range
of slopes. The overall normalization of our IMF is somewhat
arbitrary, so the key point of comparison is in the slope vs
\sigc.} \label{fig:IMF1}
\end{figure}

Given the opposing correlations of age with \fimf\ and \mst\ (Fig.~\ref{fig:downsizing}),
we might expect an anti-correlation between \fimf\ and \mst.
We do see a weak suggestion of this effect, which is a bit clearer
when considering \fimf\ versus \sigc\ (Fig.~\ref{fig:IMF1}).
This result appears inconsistent with the strong {\it positive} \fimf-\sigc\ correlation found by T+10.
However, these authors did not allow for AC in their models,
which may impact their conclusions.

As mentioned above, there are reasons to expect an IMF-age trend
for ETGs qualitatively similar to our toy model interpretation.
(Further analysis would be required to check for problems
involving FP twisting with redshift; cf.
\citealt{2006ARA&A..44..141R}.) Now considering also late-types as
relatively young, low-SFR, low-density systems, we would expect
their IMFs to be more bottom-heavy (high \fimf). Mergers of
these systems to form the younger present-day ETGs would then also
be expected to result in bottom-heavy IMFs, which is qualitatively
consistent with the trend from Fig.~\ref{fig:IMF0}, but might be
in conflict with IMF constraints in low-$z$ galaxy disks.

\section{Conclusions}\label{sec:concl}

We have continued an analysis begun in paper I of a large data-set of nearby early-type galaxies (ETGs),
combining dynamics and stellar populations to constrain the central DM content.
After having identified variations in DM as the main cause of the tilt of the fundamental plane,
we have moved on to consider various scaling relations of the DM haloes, and connections to the
star formation histories of the galaxies.

Our basic observational findings are that the central DM fraction \fDM\
within an effective radius \Re\ has a strong anti-correlation
with stellar age, and that the galaxy sizes also have an age anti-correlation.
We have constructed composite profiles of DM density with radius, finding that they are
on average cuspy, with inferred density exponents of $\sim -1.6$ near \Re.
These profiles are steeper than literature findings for spiral galaxies, and
the central DM densities of the early-types are denser overall,
suggesting that gas-rich mergers would need to produce a net halo contraction.

To further interpret the data,
we have generated a series of $\Lambda$CDM toy models, including variable contributions from
adiabatic contraction (AC).

The results from comparisons of models to data are:
\begin{itemize}
\item Models with AC fit well overall with a Kroupa IMF, while models without AC prefer a Salpeter IMF.
\item The size-age trends can explain {\it part} of the \fDM-age trends:
older galaxies show less evidence for DM because their more compact stellar centres
probe less volume of the DM halo.
\item The remaining \fDM-age trends are not easily explained by variations in halo mass or concentration, and suggest differences in baryonic effects on the DM,
in the sense that younger galaxies have undergone AC while older galaxies have not.
\item An alternative scenario is for the IMF to be less massive for older stellar populations.
\end{itemize}

There is ample scope for future insights and improvements.
We plan to further investigate
the galaxies' star formation histories in the context of theoretical mass assembly histories.
Environmental trends can be investigated, as these are expected to be important
(e.g. \citealt{Thomas+05,2006ApJ...652...71W,2010arXiv1002.0835R,2010arXiv1002.0847N,2010arXiv1003.1119L}).
Forthcoming high-quality, homogeneous, multiwavelength large surveys of low-redshift ETGs
should also be able to refute, confirm or extend the trends presented here
\citep{2009PhDT.........9G,2009arXiv0908.1904C}.
Finally, the gold standard for probing galaxy mass profiles is extended kinematics data along
with detailed dynamical modelling---both to provide more leverage on the DM independently of
the stellar mass, and to sift individual galaxies for the presence of a DM core or cusp
(e.g. \citealt{2007MNRAS.382..657T,2009MNRAS.395...76D,2009MNRAS.398..561W,2009PhDT.........7F}).

Even if some of our current conclusions turn out to be completely wrong,
we hope to have introduced a useful framework for
interpreting mass results for large data sets of ETGs over cosmic time.
The DM constraints are part of a spectrum of clues that can ultimately be
combined to pin down the modes of ETG formation (e.g. \citealt{2008MNRAS.384...94C}).
Other avenues with considerable promise include
central rotation \citep{2009MNRAS.397.1202J}, extra light \citep{2010MNRAS.402..985H}
and orbital structure \citep{2008ApJ...685..897B}, as well as
halo rotation \citep{2010arXiv1001.0799H} and metallicity gradients \citep{2009MNRAS.398..561W,2009MNRAS.400.2135F},
and globular cluster constraints \citep{2007AJ....134.1403R,2009ApJ...691...83S,2010MNRAS.401L..58B}.

\section*{Acknowledgements}

We thank Stacy McGaugh for providing his data tables in electronic form, and
Surhud More and Frank van den Bosch for sharing their paper in advance of publication.
We thank Nate Bastian, Michele Cappellari, J\"urg Diemand, Aaron Dutton,
Gianfranco Gentile, Alister Graham, Brad Holden, Mike Hudson, Fill Humphrey, David Koo,
Claire Lackner, Stacy McGaugh, Joel Primack, Tommaso Treu and Mike Williams for helpful discussions,
and the anonymous referee for constructive comments.
AJR was supported by National Science Foundation Grants AST-0507729, AST-0808099, and AST-0909237.
CT was supported by the Swiss National Science Foundation and by a grant
from the project Mecenass, funded by the Compagnia di San Paolo.

\appendix

\section{Investigating negative dark matter fractions}\label{sec:unphys}

Our derivation of \fDM\ from estimates of \Yst\ and \Ydyn\ yields a number of
cases where galaxies have an unphysical $\fDM < 0$.
These cases comprise 2\%, 7\% and 25\% of the sample for
the Chabrier, Kroupa and Salpeter IMFs, respectively.
Here we investigate whether these fractions could be compatible with simple observational
scatter in \Yst\ and \Ydyn.

We have estimated the uncertainties in \Yst\ to be $\sim$~15\% using Monte Carlo
simulations of SED fitting.
For \Ydyn, we may generically consider it as derived via a virial relation:
\begin{equation}
\Ydyn = \frac{K \sigma_{\rm eff}^2 \Re}{G L} ,
\end{equation}
where $K$ is a virial coefficient, $G$ is the gravitational constant, and $L$ is the luminosity
(see C+06 equation 19).
In this context, the value of \Re\ is arbitrary, and the uncertainty in the physical units
$\Re/L$ is dominated by the $\sim$~10\% distance uncertainty.
To estimate $\sigma_{\rm eff}$ we first consider the central \sigc\ with a measurement error of $\sim$~5\%,
and extrapolate this to the value averaged over \Re---an exercise with a $\sim$~5\% uncertainty
(C+06; we consider systematics separately).
The value of $K$ was found by C+06 to have a $\sim$~15\% scatter.
The net uncertainty in \Ydyn\ is then $\sim$~23\%,
and the uncertainty on \fDM\ varies from $\sim$~0.1 to $\sim$~0.2 for
high to low \fDM\ values.

\begin{figure}
\hspace{-0.3cm} \epsfig{file=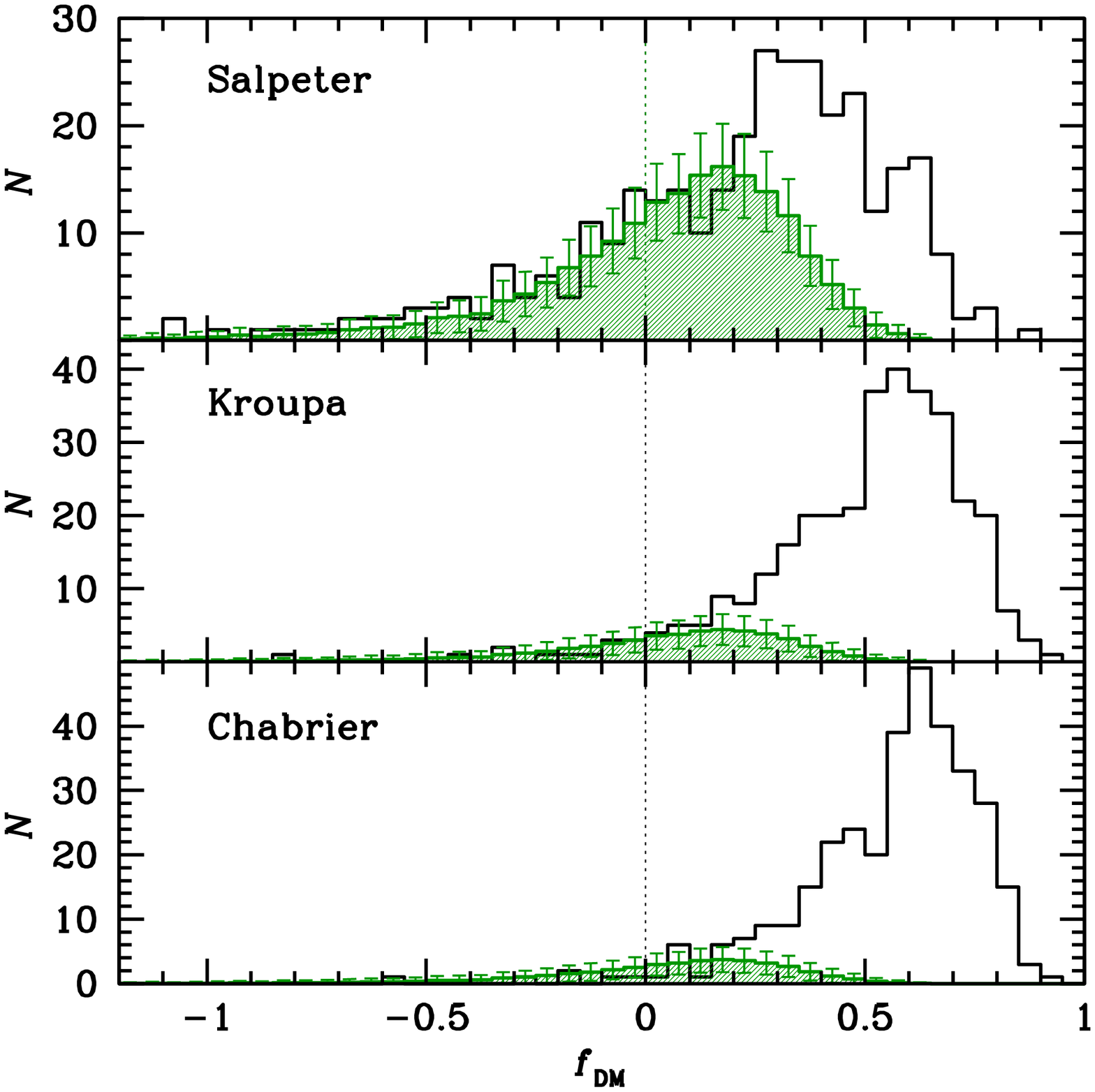,
width=0.51\textwidth} \caption{Distributions of DM fraction for early-type galaxies.
The open histograms are the observational results, with a different IMF for each panel
(as labelled).  The shaded green histograms are random realizations of galaxy subsamples
having intrinsic $\fDM=0.1$, with error bars illustrating the approximate
Poissonian uncertainties.
See text for further details.
}\label{fig:neg}
\end{figure}

We begin by assuming some fraction of the galaxy sample have true $\fDM \sim 0.1$,
which we find is the lowest plausible value in a $\Lambda$CDM context (Section~\ref{sec:lcdm1}).
Using the uncertainties discussed above, we generate a random sample of \fDM\ measurements and scale
a histogram of their frequency to reproduce the number of $\fDM < 0$ observations while
not exceeding any of the $\fDM > 0$ frequencies (Fig.~\ref{fig:neg}).
For the Chabrier and Kroupa IMFs, this scaling process implies that
$\sim$~15\%
of the sample have $\fDM \sim 0.1$.
For the Salpeter IMF on the other hand, it is difficult to find a scaling that reproduces
the negative-\fDM\ tail without violating
the observed distribution to higher values; the best fit has $\sim$~55\% with $\fDM \sim 0.1$.

This difficulty with Salpeter would be eased if a true $\fDM=0$ is assumed for half the galaxies,
if there were systematic errors at the level of $\sim$~15\%, or if
the random \Yst\ measurement errors were
actually at the level of $\sim$~25\%, or were non-Gaussian.
We would not at present rule out any of these possibilities and so cannot
categorically exclude a universal Salpeter IMF.
Recalibrating our $\Ydyn$ and $\Yst$ estimates to external results (as discussed
in paper I Appendix A) would on the other hand make the situation more problematic for
Salpeter by increasing the number of objects with $\fDM < 0$.

\section{Cross-checks on age dependencies}\label{sec:app}

Here we carry out various tests on the robustness of the DM-age trends found in
Section~\ref{sec:fdm_sfe}.
In Section~\ref{sec:appfDM} we compare our \fDM\ results to other
literature results, and
in Section~\ref{sec:largeR} we examine the implications of
results at larger radii.
We explore systematic effects in our stellar populations models
in Section~\ref{sec:modsys}.

\subsection{Central DM content}\label{sec:appfDM}

We check here whether our results on central \fDM\ and age
are consistent with other
results in the literature.  First we consider the SDSS-based
analysis of SB09.  They estimated central
dynamical and stellar masses for a large sample of ETGs at $z ~\sim$~0.1--0.2
using somewhat different techniques to ours in paper I.
After correcting the galaxy luminosities to a common age, they
found total \ML\ trends that depend systematically on age (their Fig.~2).
Using the information in \citet{2009MNRAS.395.1491B} to convert these corrected
luminosities to stellar masses, we convert their \ML\ values to
$M_{\rm dyn}/M_\star$ and then to \fDM.
It also appears that they assumed a Chabrier IMF, so we further convert their
results to a Kroupa IMF for comparison with our default models and results.

\begin{figure}
\hspace{-0.3cm} \epsfig{file=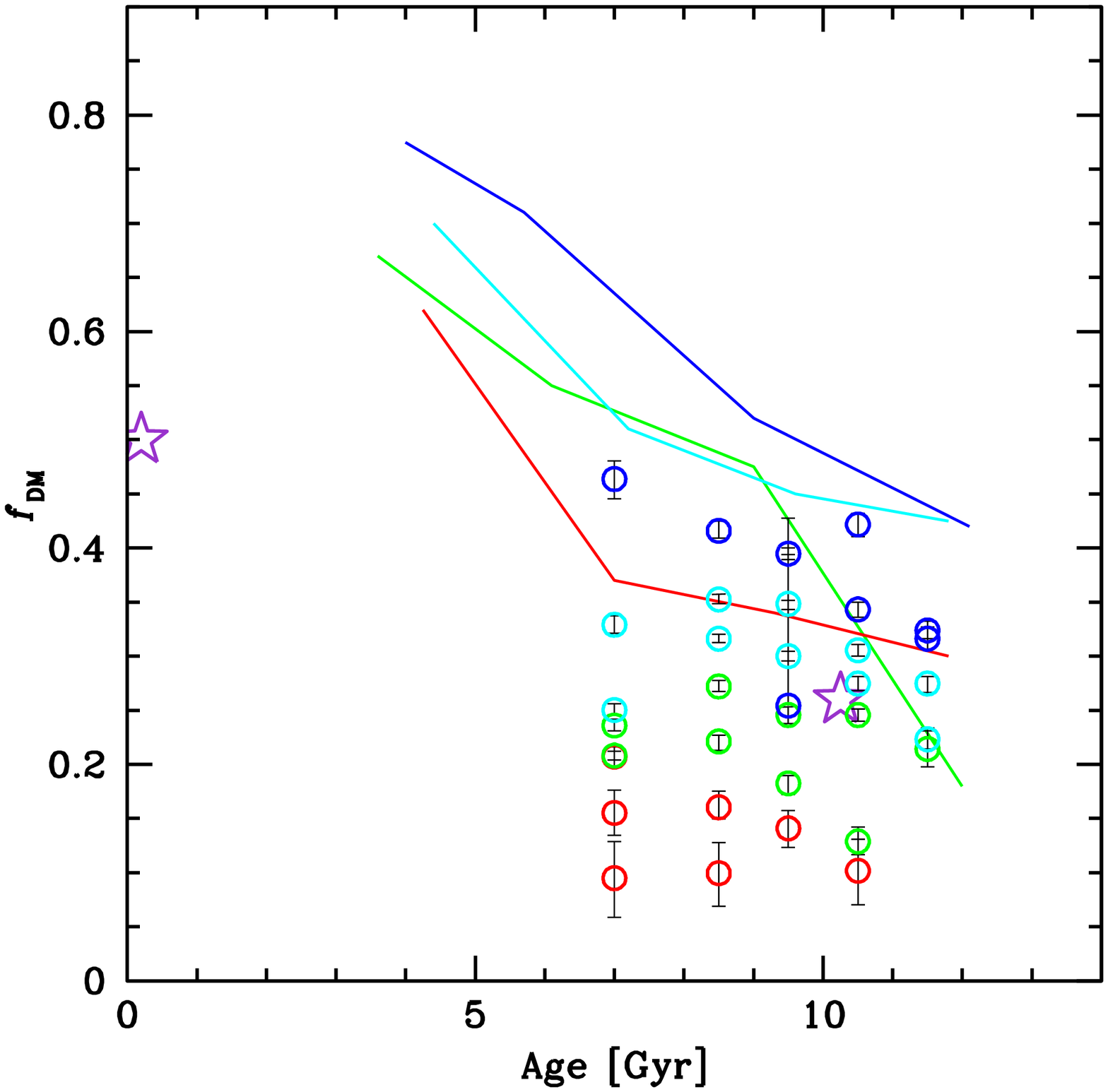,
width=0.51\textwidth} \caption{DM fraction with age of $z\sim$~0.1--0.2 ETGs, in bins of
stellar mass, from SB09;
the data have been converted from using a Chabrier IMF to a Kroupa IMF.
The colours show the same mass bins as in Fig.~\ref{fig:LCDM_fdmage}.
For comparison, the lines show our median results in each mass bin.
Also shown as large stars are typical results from low- and high-$z$
ETG samples from C+06 and \citet{2009ApJ...704L..34C}.
The ``ages'' plotted in this case actually correspond to the redshifts of the observations,
but we do not have enough information to correct these to $z=0$ ages (which will
qualitatively be higher than the plotted points).
}\label{fig:SB09}
\end{figure}

The results of this exercise are plotted in Fig.~\ref{fig:SB09}:
there are indications of an anti-correlation between \fDM\ and age,
particularly for the higher mass bins.
For comparison, our results are also plotted;
other than a 30\% overall offset in \ML\ whose origin is unclear,
the results are generally consistent.
The SB09 data do suggest somewhat shallower
correlations that might be explained by a pure \Re-age effect
(as stated by these authors) but they do not cover a large enough
range in age to be sure.

\citet{2009ApJ...698.1590G} carried out a different analysis of quiescent ETGs in the SDSS.
They again analyzed the central dynamical and stellar masses separately,
and mapped various stellar populations parameters on approximate slices
of the FP defined by \sigc\ and \Re.
Although these results do imply a systematic correlation between size and
age at fixed mass, if one considers a single \sigc-\Re\ grid-point,
then one can control for size and mass dependencies and look for any
residual correlations in the perpendicular direction of {\it surface brightness}.

These authors found a negative correlation between age and surface brightness,
which because dynamical mass is approximately constant implies that there is
a positive correlation between age and overall \ML.
They mentioned that in a forthcoming paper they will find that stellar populations
effects are not enough to explain this age-\ML\ correlation,
which therefore implies a net positive age-\fDM\ correlation.
This result is {\it opposite} to the trend that we find for a residual negative
age-\fDM\ correlation after the age-\Re\ correlations are accounted for.
It remains to be seen exactly how and why our results differ.

\citet{2010arXiv1001.2965B} Fig.~4 provides an intriguing compilation of comparisons between \Ydyn\ and
\Yst\ at $z=0$ and $z\sim2$ for ETGs from C+06 and \citet{2009ApJ...704L..34C}.
While there are many issues in making such a direct comparison between the
two galaxy samples
the apparent trend does agree qualitatively with our results,
although we cannot evaluate the potential residuals from a size-driven trend
(Fig.~\ref{fig:SB09}).

\begin{figure}
\hspace{-0.6cm} \epsfig{file=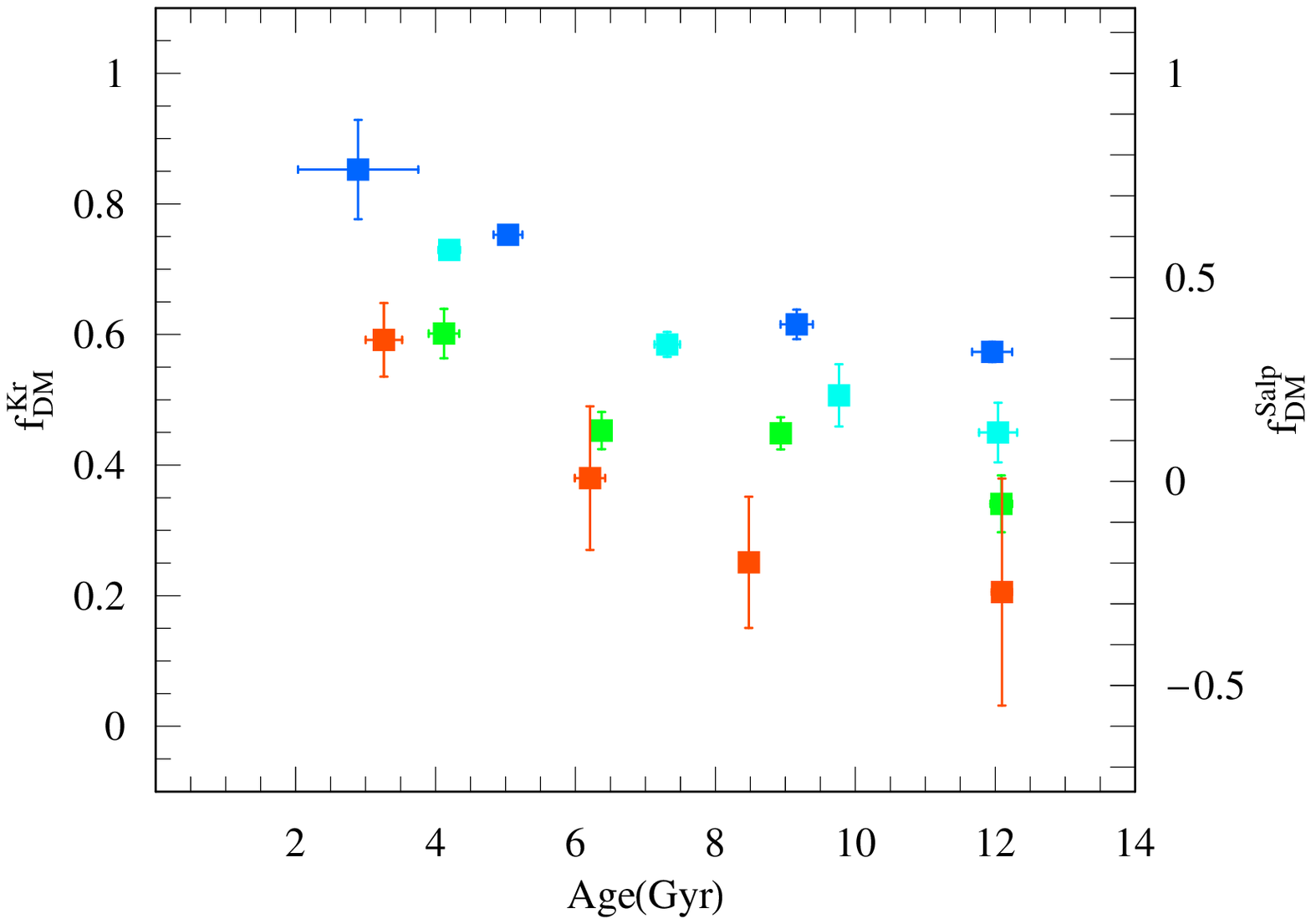, width=0.52\textwidth}
\caption{Our results for central DM fraction versus age, in bins of
constant \sigc\ ($\sim$~120, 180, 250, 300 \kms).
}
\label{fig:fDMsig0}
\end{figure}

Since \citet{2009ApJ...698.1590G} among others emphasize the primary importance of
\sigc\ (rather than \mst) in correlating with stellar populations parameters,
we show in Fig.~\ref{fig:fDMsig0} our results for \fDM-age in bins of constant \sigc.
The trends are similar to those found in bins of constant \mst.
However, we do not use \sigc\ in general in this paper because it is less
straightforward to incorporate in the toy models, and because \fDM\ depends on
a dynamical mass determination that is intrinsically correlated with \sigc.

\subsection{Large radius DM content}\label{sec:largeR}

\begin{figure*}
\epsfig{file=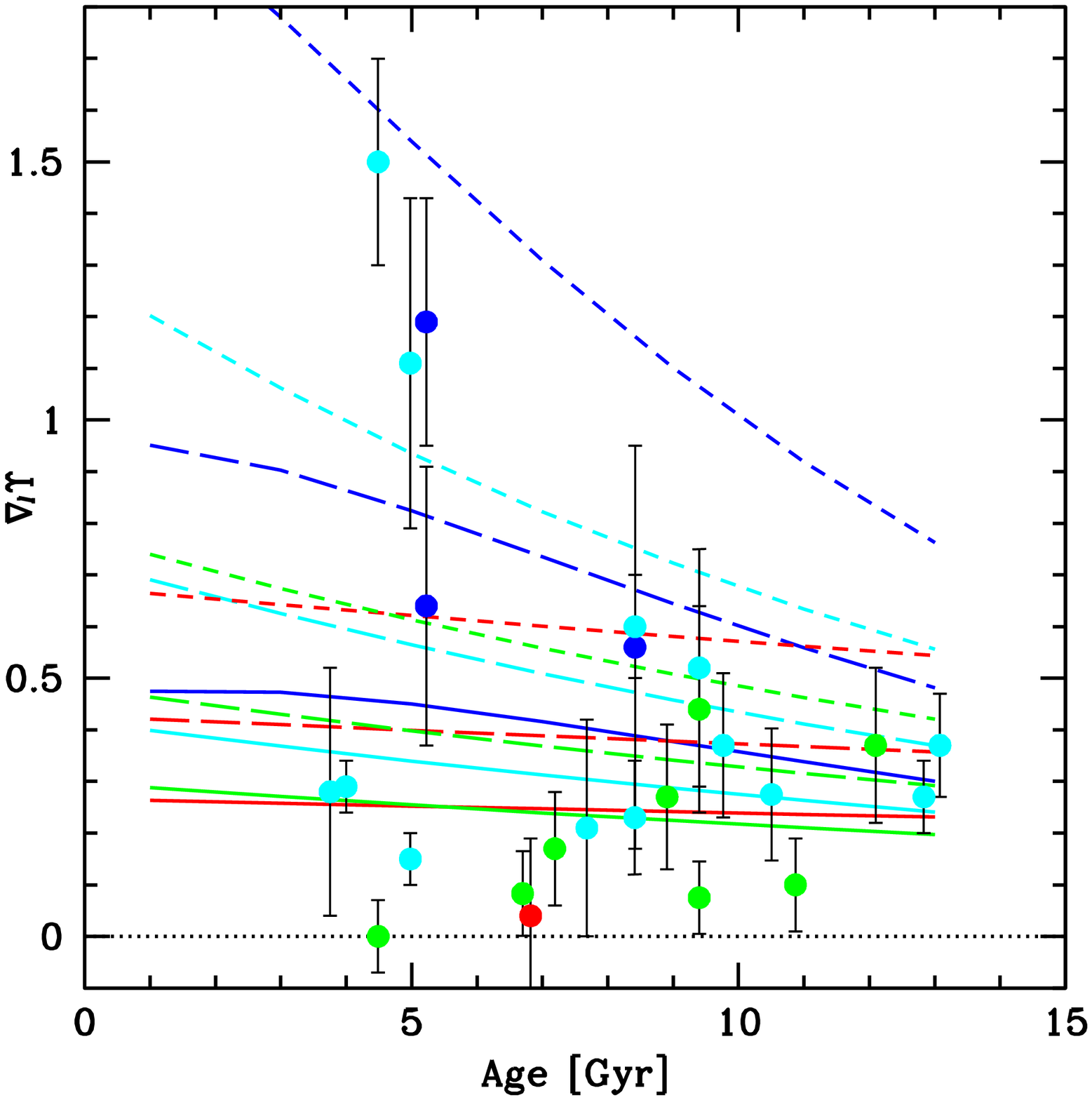, width=0.49\textwidth}
\epsfig{file=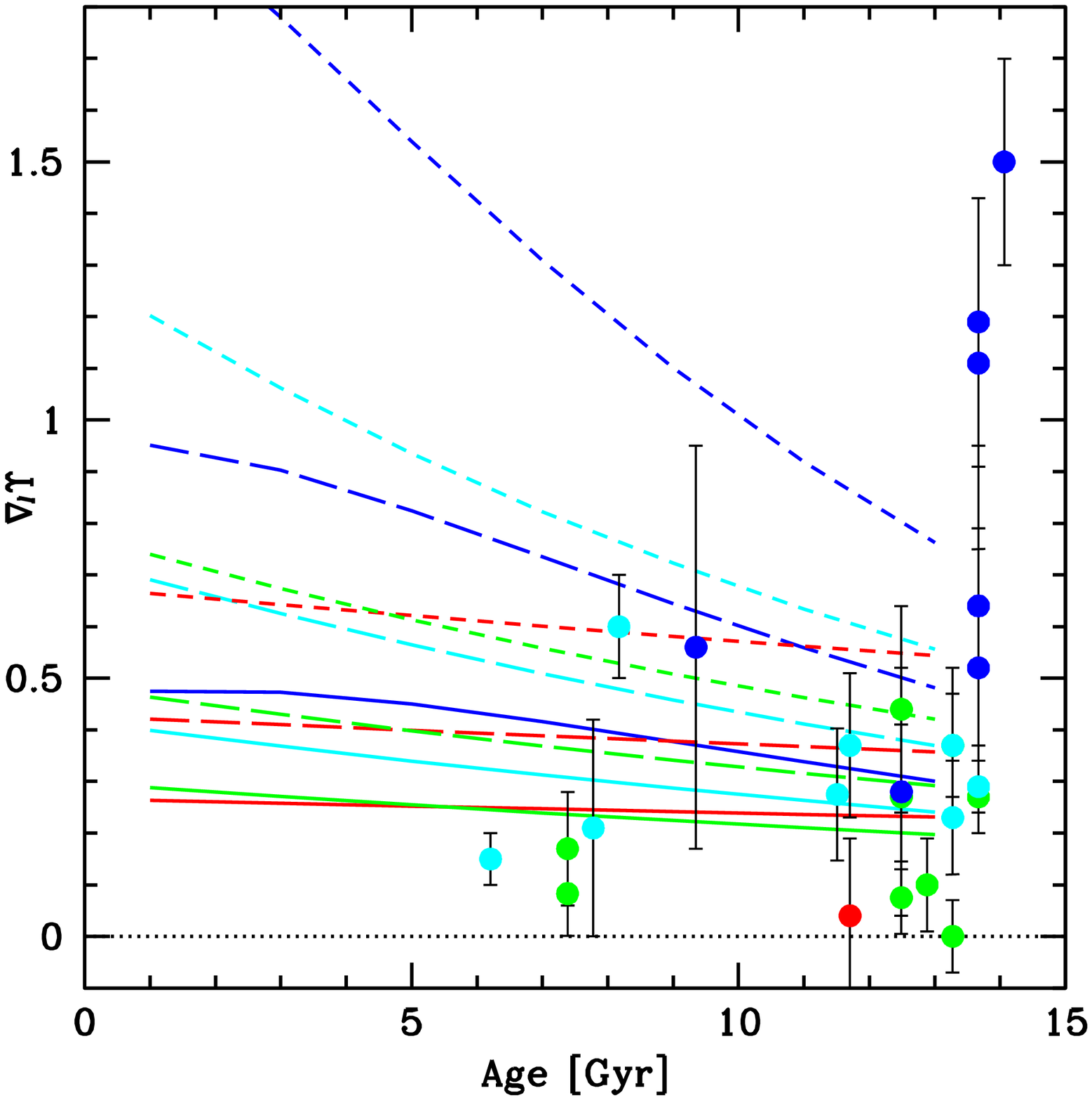, width=0.49\textwidth}\\
\epsfig{file=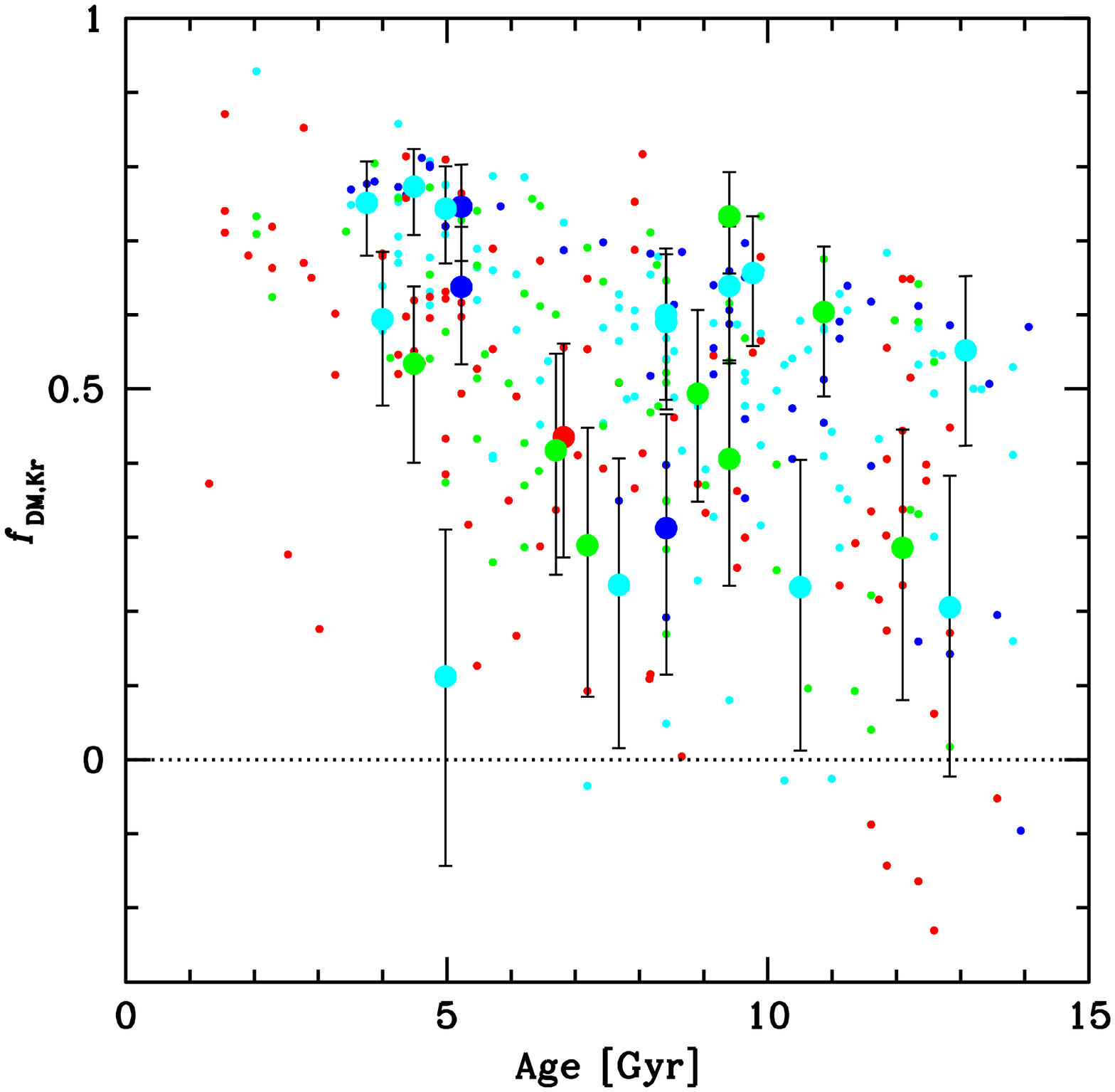, width=0.49\textwidth}
\epsfig{file=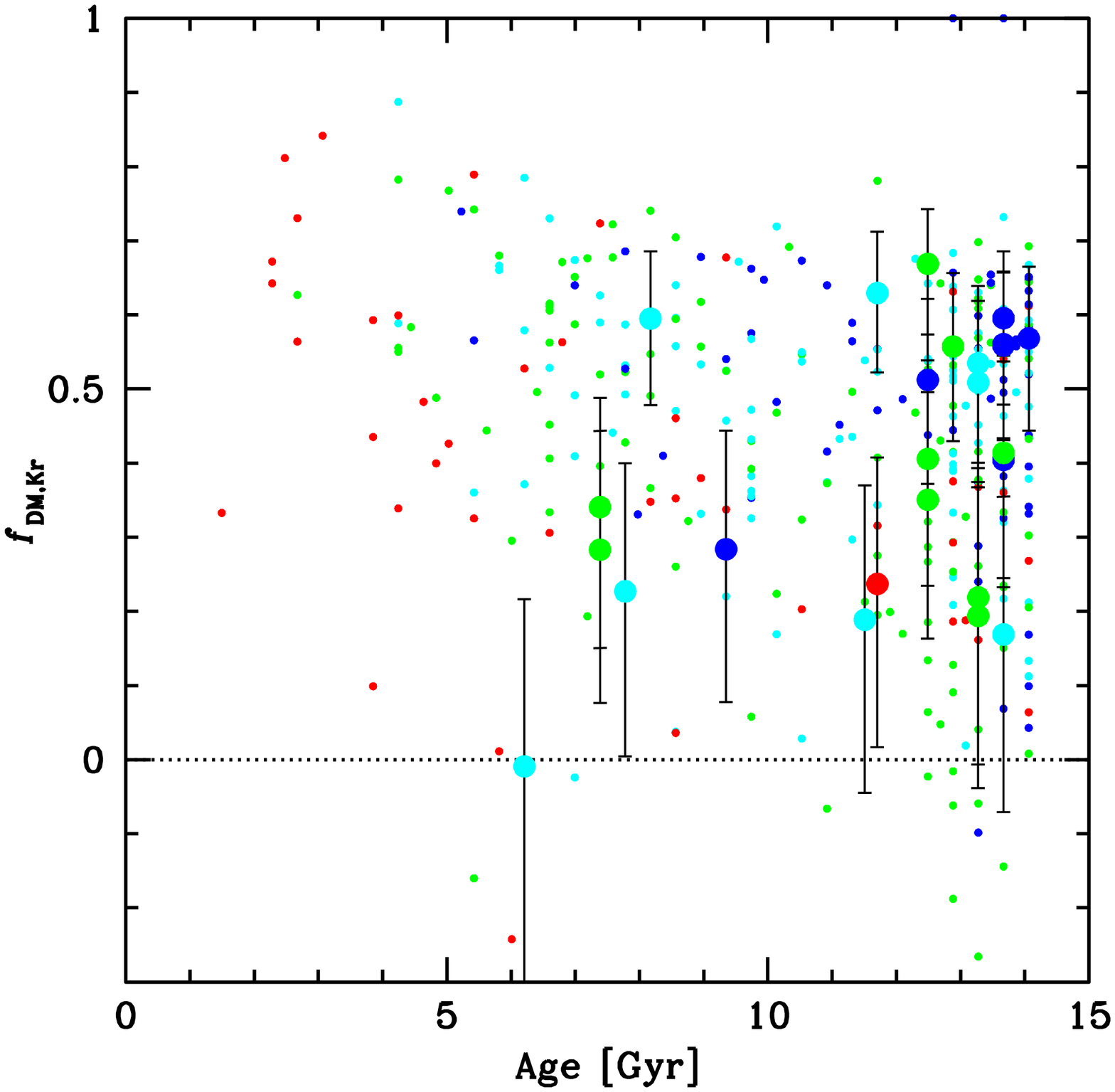, width=0.49\textwidth} \caption{ DM-age
trends in a subsample of galaxies with large-radius dynamical
tracers. The {\it top row} shows the ``mass-to-light ratio radial
gradient'', which as an observational parameter is based on
dynamics only and is independent of the IMF. Colourised model
curves and mass bins are as in Fig.~\ref{fig:LCDM_fdmage}. The
models are for the case of Kroupa IMF using a G+04 AC recipe, with
line-styles showing different \esf\ values (note that AC has only
a small impact on these model predictions since it is a process
that acts strongly at small radii). The {\it bottom row} shows the
DM fraction within \Re, with the small symbols for our full galaxy
sample (compare Fig.~\ref{fig:fDM_age}) and large symbols for the
subsample with large-radius results. The {\it left panels} show
results based on our general stellar populations models with
metallicity left as a free parameter, and the {\it right panels}
show the results with metallicity fixed to Solar ($Z=0.02$). }
\label{fig:grad}
\end{figure*}

Next we consider DM conclusions from dynamical studies
extending to large galactocentric radii, quantified as the \ML\ gradient
parameter introduced in \citet{Nap05}.
The data-set of 25 galaxies is from paper I (Fig.~C2, with an update on NGC~4374 from
Napolitano et al., in prep), where we confirmed that the gradient
correlated well with \fDM\ from the central regions.
We show some of our toy model predictions from this paper in
Fig.~\ref{fig:grad} (upper left panel).
In the absence of any residual DM-age trends, the gradient is expected to decrease with age,
since it is defined relative to the galaxy \Re, which in turn decreases with age.

At first glance, the data appear to support this expectation, with the residuals
unclear because of the scatter.
However, closer examination of the data-sets plotted raises a red flag: the four ``young'',
DM-dominated galaxies are all well-known nearby massive group- and cluster-central
ellipticals that are normally thought to have very old stellar populations (averaged
over \Re), {\it not} with ages $\sim$~4--5 Gyr (NGC~1407, M49, M87, NGC~5846;
e.g. C+06; \citealt{2007MNRAS.377..759S,2008MNRAS.385..675S}).

It turns out that these galaxies were all fitted with an unrealistic super-solar metallicity
($Z=0.05$) which is probably a reflection of the fundamental age-metallicity degeneracy in
stellar populations analyses.
We re-ran our analysis with all the metallicities fixed to
$Z=0.02$, which for the four problematic galaxies yielded more credible ages of
14~Gyr\footnote{For galaxies with a previous best fit of $Z=0.02$, the age and \Yst{}
results with fixed $Z$ are {\it not} exactly the same as before.  This is because our
procedure does not involve a simple best fit, but rather a Monte Carlo approach using the
median of a distribution of best-fit values: see paper I.}.
The revised $M/L$-gradient results are shown in the top right panel of Fig.~\ref{fig:grad},
where it now appears that the DM content may {\it increase} with age.
The first thing to keep in mind when considering this apparent inconsistency with our
main \fDM-age result is that the DM content within \Re{} and within $\sim$~5~\Re{}
may very
well {\it not} be tightly correlated.  The central DM content may be less a reflection of the
overall DM content and more closely related to the details of the baryonic-DM interplay
at the centres of haloes.

The main effect however appears to be small number statistics coupled with selection effects.
In the bottom panels of Fig.~\ref{fig:grad} we show the \fDM\ results for both
metallicity assumptions.  The stellar $M/L$ is not very sensitive to the age-metallicity
degeneracy, and in fact is affected in a way that roughly parallels the overall
\fDM-age trend: higher age and lower $Z$ yield higher $M/L_\star$ and lower \fDM.
Thus the general trend for our full sample is not qualitatively affected by changing $Z$,
but quantitatively shifts to larger ages overall (see also Fig.~\ref{fig:models}).

The subsample of galaxies that have large-radius data is not so fortunate.
When fixing $Z=0.02$, their \fDM-age results buck the overall trend, just
as seen from the large-radius results.  This sample has only one ``young'' object
($\lsim$~7 Gyr), and the old objects are dominated by systems like M87 which are
known to reside at the centres of massive groups and clusters, and appear to comprise
the high-\fDM{} tail of the overall distribution.
Such galaxies are some of the first targets of large-radius dynamical studies because of their rich
supply of mass tracers such as globular clusters and planetary nebulae.
The price to pay from selecting such systems
is that they could provide a very biassed view of the
Universe---as now appears to be the case when considering central DM content.

This selection effect may explain the curious DM halo mass-concentration trends found
by N+09, and demonstrates the importance of constructing an
unbiassed galaxy sample.
For now, the DM-age implications from large-radius tracers are totally inconclusive,
and will require completion of a large systematic survey to make any progress.
In the meantime, valuable spot-checks could be provided by studying the large-radius
DM content of a few galaxies that help drive the apparent \fDM-age relation, e.g.
young DM-rich systems like NGC~3626 or old DM-poor ones like NGC~7454.

\clearpage

\clearpage

\subsection{Modelling systematics}\label{sec:modsys}

We next consider whether systematic uncertainties in our stellar populations
modelling could be affecting our DM inferences.
We have already explored this issue for the \fDM-mass relations in paper I;
here we consider \fDM-age.
Age and stellar \ML\ are both derived from the same models,
and a positive error in age would correlate with a positive error in stellar \ML\,
which would produce a {\it negative} error in \fDM\ and thus mimic
an anti-correlated \fDM-age trend solely because of the correlated errors.

\begin{figure}
\hspace{-0.8cm} \epsfig{file=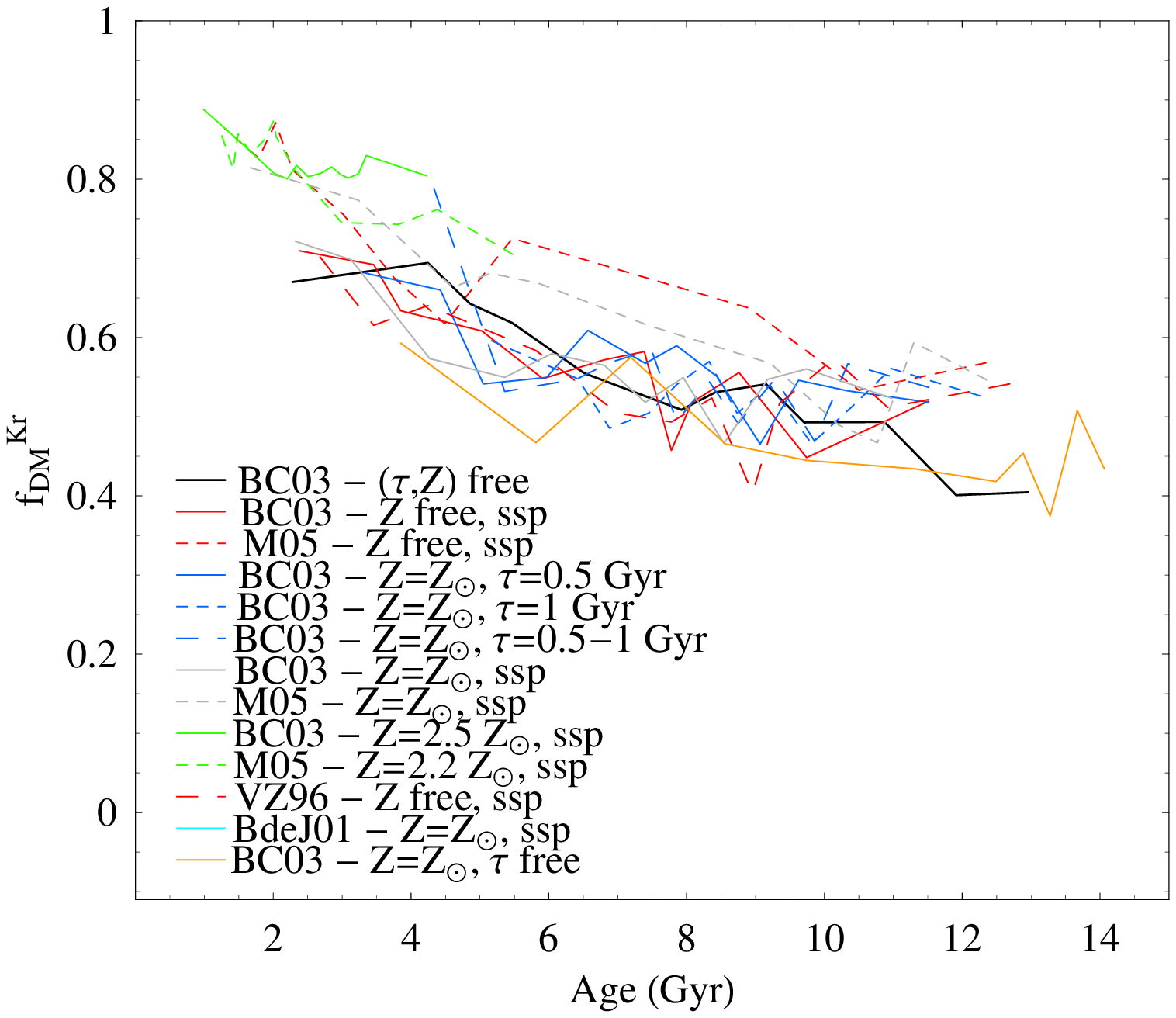, width=0.51\textwidth}
\caption{Trend of DM fraction with age, for different stellar
populations assumptions.  The solid black line is our standard
model, and see Fig.~A2 of paper I for explanations of the other
line styles. An additional model not included here is from
\citet{2009ApJ...699..486C}, but for the wavelengths and ages used
here, this would be equivalent to BC03 (see their Fig.~10). We
have also tested delayed-exponential and truncated (step-function)
SFH models, which are not shown here but again do not
substantially change the trends above.}\label{fig:models}
\end{figure}

To see if such a systematic could account for our observational result,
we experiment with different stellar populations assumptions, showing the
results in Fig.~\ref{fig:models}.
The only cases where the \fDM-age trend is appreciably different
from our standard estimate is when we unrealistically fix $Z=0.05$ (green lines).

\begin{figure}
\hspace{-0.5cm} \epsfig{file=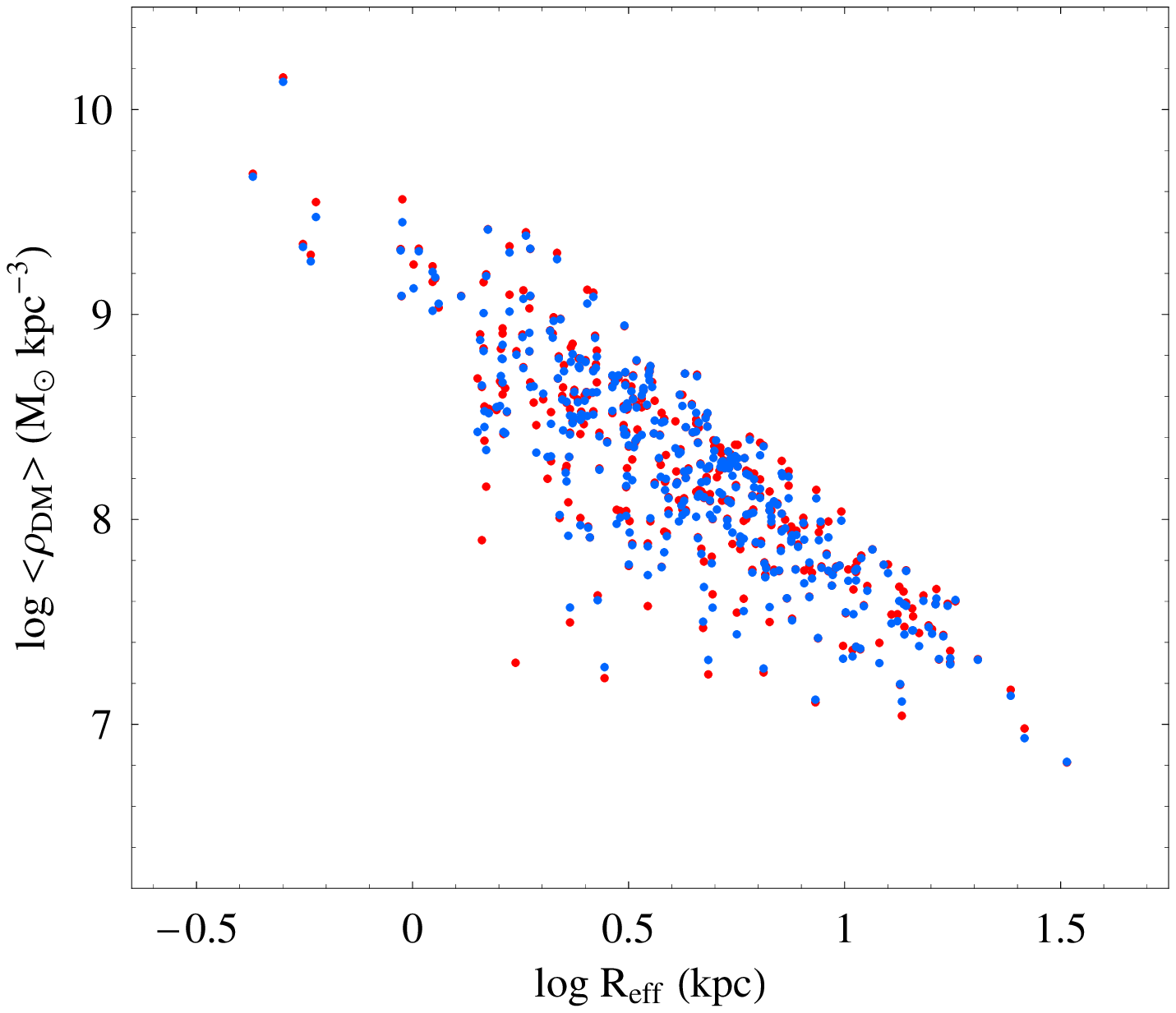, width=0.51\textwidth}
\caption{Central DM density versus effective radius for our full
ETG galaxy sample.  Red points show results using our standard
assumptions in the stellar populations models, and blue shows the
case where metallicity is fixed to solar. }\label{fig:modrhoDM}
\end{figure}

Given the problem identified in Section~\ref{sec:largeR} where allowing
the metallicity to vary freely can result in too many ``young'' galaxies,
we consider in particular our results when we fix $Z=\Zsun$.
The implications for \fDM\ were shown in Figs.~\ref{fig:grad} (right panels) and
\ref{fig:models} (solid orange line),
and we also show the overall \rhoDM-\Re\ trend in Fig.~\ref{fig:modrhoDM}.
The results are similar to our standard model.

As a reminder, in paper I (Appendix A) we compared independent estimates of \Ydyn\ and
\Yst\ to C+06 for galaxies in common.
Our \Yst\ values for the same IMF were $\sim$~20\% lower, and our \Ydyn\ values were higher
(by $\sim$~10\% for the brighter galaxies and $\sim$~30\% for the fainter galaxies).
Thus our \fDM\ values are systematically higher, becoming more discrepant for fainter galaxies
(up to $\Delta \fDM\ \sim 0.25$).
The C+06 results on their own may imply consistency with no-AC models for the fainter
galaxies, and strong AC for the brighter ones \citep{2006EAS....20..119R},
which would be in better agreement with the results of N+09.
We have also checked the \fDM-age trends using a recalibration as discussed in paper I,
and found that for the fainter galaxies, the anti-correlation becomes somewhat {\it steeper},
so our basic \fDM-age result does not go away.

An issue not discussed in paper I is the potential effect of AGN emission on the
observed galaxy colours and thus on the \Yst\ and age inferences.
The AGN colours would generally mimic a stellar population of $\sim$~3~Gyr age
and skew the inferences toward younger ages and higher \fDM.
Matching the SDSS DR4 ETG catalogue with the AGN catalogue of \citet{2003MNRAS.346.1055K},
we find that $\sim$~15\% of the galaxies in our sample's mass range have strong AGNs
(Type II with strong [O III] emission; see also \citealt{2009ApJ...693..486G}).
However, based on the AGN study of \citet{1999MNRAS.303..173S}, we estimate that the contaminant light
would account for only $\sim$~1--2\% of the total light within \Re, and would
bias our results by at most 5\%.


\begin{thebibliography}{}

\bibitem[\protect\citeauthoryear{Abadi et al.}{2009}]{2009arXiv0902.2477A} Abadi, M.~G., Navarro, J.~F., Fardal, M., Babul, A., \& Steinmetz, M.\ 2009, arXiv:0902.2477

\bibitem[Agertz et al.(2009)]{2009MNRAS.397L..64A} Agertz, O., Teyssier, R., \& Moore, B.\ 2009, MNRAS, 397, L64

\bibitem[Allanson et al.(2009)]{2009ApJ...702.1275A} Allanson, S.~P., Hudson, M.~J., Smith, R.~J., \& Lucey, J.~R.\ 2009, ApJ, 702, 1275

\bibitem[Angus et al.(2008)]{2008MNRAS.387.1470A} Angus, G.~W., Famaey, B., \& Buote, D.~A.\ 2008, MNRAS, 387, 1470

\bibitem[Auger et al.(2009)]{2009ApJ...705.1099A} Auger, M.~W., Treu, T., Bolton, A.~S., Gavazzi, R., Koopmans, L.~V.~E., Marshall, P.~J., Bundy, K., \& Moustakas, L.~A.\ 2009, ApJ, 705, 1099

\bibitem[Barnab\`e et al.(2010)]{2010arXiv1002.1083B} Barnab\`e, M., Auger, M.~W., Treu, T., Koopmans, L., Bolton, A.~S., Czoske, O., \& Gavazzi, R.\ 2010, MNRAS, submitted, arXiv:1002.1083

\bibitem[Bastian et al.(2010)]{2010arXiv1001.2965B} Bastian, N., Covey, K.~R., \& Meyer, M.~R.\ 2010, ARAA, in press, arXiv:1001.2965

\bibitem[Bekki \& Shioya(1998)]{1998ApJ...497..108B} Bekki, K., \& Shioya, Y.\ 1998, ApJ, 497, 108


\bibitem[Bekki(2010)]{2010MNRAS.401L..58B} Bekki, K.\ 2010, MNRAS, 401, L58

\bibitem[Bernardi(2009)]{2009MNRAS.395.1491B} Bernardi, M.\ 2009, MNRAS, 395, 1491

\bibitem[Bezanson et al.(2009)]{2009ApJ...697.1290B} Bezanson, R., van Dokkum, P.~G., Tal, T., Marchesini, D., Kriek, M., Franx, M., \& Coppi, P.\ 2009, ApJ, 697, 1290

\bibitem[Binney et al.(2001)]{2001MNRAS.321..471B} Binney, J., Gerhard, O., \& Silk, J.\ 2001, MNRAS, 321, 471

\bibitem[\protect\citeauthoryear{Blumenthal et al.}{1986}]{Blumenthal+86} Blumenthal, G.~R., Faber, S.~M., Flores, R., \& Primack, J.~R.\ 1986, ApJ, 301, 27 (B+86)

\bibitem[Bolton et al.(2007)]{2007ApJ...665L.105B} Bolton, A.~S., Burles, S., Treu, T., Koopmans, L.~V.~E., \& Moustakas, L.~A.\ 2007, ApJ, 665, L105

\bibitem[Bolton et al.(2008)]{2008ApJ...684..248B} Bolton, A.~S., Treu, T., Koopmans, L.~V.~E., Gavazzi, R., Moustakas, L.~A., Burles, S., Schlegel, D.~J., \& Wayth, R.\ 2008, ApJ, 684, 248

\bibitem[Boyarsky et al.(2009)]{2009arXiv0911.3396B} Boyarsky, A., Neronov, A., Ruchayskiy, O., \& Tkachev, I.\ 2009, arXiv:0911.3396

\bibitem[Boylan-Kolchin \& Ma(2004)]{2004MNRAS.349.1117B} Boylan-Kolchin, M., \& Ma, C.-P.\ 2004, MNRAS, 349, 1117

\bibitem[Boylan-Kolchin et al.(2005)]{2005MNRAS.362..184B} Boylan-Kolchin, M., Ma, C.-P., \& Quataert, E.\ 2005, MNRAS, 362, 184

\bibitem[Boylan-Kolchin et al.(2006)]{2006MNRAS.369.1081B} Boylan-Kolchin, M., Ma, C.-P., \& Quataert, E.\ 2006, MNRAS, 369, 1081

\bibitem[Brown et al.(2008)]{2008ApJ...682..937B} Brown, M.~J.~I., et al.\ 2008, ApJ, 682, 937

\bibitem[Brownstein \& Moffat(2006)]{2006ApJ...636..721B} Brownstein, J.~R., \& Moffat, J.~W.\ 2006, ApJ, 636, 721

\bibitem[\protect\citeauthoryear{Bruzual \& Charlot}{2003}]{BC03} Bruzual, A. G. \& Charlot, S. 2003, MNRAS, 344, 1000

\bibitem[Bullock et al.(2001)]{2001MNRAS.321..559B} Bullock, J.~S., Kolatt, T.~S., Sigad, Y., Somerville, R.~S., Kravtsov, A.~V., Klypin, A.~A., Primack, J.~R., \& Dekel, A.\ 2001, MNRAS, 321, 559

\bibitem[Buote et al.(2007)]{2007ApJ...664..123B} Buote, D.~A., Gastaldello, F., Humphrey, P.~J., Zappacosta, L., Bullock, J.~S., Brighenti, F., \& Mathews, W.~G.\ 2007, ApJ, 664, 123

\bibitem[Burkert(1995)]{1995ApJ...447L..25B} Burkert, A.\ 1995, ApJ, 447, L25

\bibitem[Burkert et al.(2008)]{2008ApJ...685..897B} Burkert, A., Naab, T., Johansson, P.~H., \& Jesseit, R.\ 2008, ApJ, 685, 897

\bibitem[Burkert et al.(2010)]{2009arXiv0907.4777B} Burkert, A., et al.\ 2010, ApJ, submitted, arXiv:0907.4777

\bibitem[Calura \& Menci(2009)]{2009MNRAS.400.1347C} Calura, F., \& Menci, N.\ 2009, MNRAS, 400, 1347

\bibitem[\protect\citeauthoryear{Cappellari et al.}{2006}]{Cappellari06} Cappellari, M. et al. 2006, MNRAS, 366, 1126 (C+06)

\bibitem[Cappellari et al.(2009)]{2009ApJ...704L..34C} Cappellari, M., et al.\ 2009, ApJ, 704, L34

\bibitem[Cappellari et al.(2010)]{2009arXiv0908.1904C} Cappellari, M., et al.\ 2010, in Dark Matter in Early-Type Galaxies, eds. L.V.E. Koopmans \& T. Treu, in press, arXiv:0908.1904


\bibitem[Cardone, Angus, \& Tortora(2010)]{CT10} Cardone, V.~F. \& Tortora, C. 2010, MNRAS, submitted

\bibitem[Cardone et al.(2010)]{C+10} Cardone, V.~F., et al., 2010, MNRAS, submitted

\bibitem[Carlberg(1986)]{1986ApJ...310..593C} Carlberg, R.~G.\ 1986, ApJ, 310, 593

\bibitem[\protect\citeauthoryear{Chabrier}{2001}]{Chabrier01} Chabrier, G. 2001, ApJ, 554, 1274

\bibitem[\protect\citeauthoryear{Chabrier}{2002}]{Chabrier02} Chabrier, G. 2001, ApJ, 567, 304

\bibitem[\protect\citeauthoryear{Chabrier}{2003}]{Chabrier03} Chabrier, G. 2003, PASP, 115, 763

\bibitem[Coccato et al.(2009)]{2009MNRAS.394.1249C} Coccato, L., et al.\ 2009, MNRAS, 394, 1249

\bibitem[Conroy et al.(2008)]{2008ApJ...679.1192C} Conroy, C., Shapley, A.~E., Tinker, J.~L., Santos, M.~R., \& Lemson, G.\ 2008, ApJ, 679, 1192

\bibitem[Conroy \& Wechsler(2009)]{2009ApJ...696..620C} Conroy, C., \& Wechsler, R.~H.\ 2009, ApJ, 696, 620

\bibitem[Conroy et al.(2009)]{2009ApJ...699..486C} Conroy, C., Gunn, J.~E., \& White, M.\ 2009, ApJ, 699, 486

\bibitem[Cooper et al.(2010)]{2009arXiv0910.3211C} Cooper, A.~P., et al.\ 2010, arXiv:0910.3211

\bibitem[Covington(2008)]{2008PhDT........13C} Covington, M.~D.\ 2008, Ph.D.~Thesis, Univ. California, Santa Cruz

\bibitem[Covington et al.(2008)]{2008MNRAS.384...94C} Covington, M., Dekel, A., Cox, T.~J., Jonsson, P., \& Primack, J.~R.\ 2008, MNRAS, 384, 94

\bibitem[Daddi et al.(2005)]{2005ApJ...626..680D} Daddi, E., et al.\ 2005, ApJ, 626, 680

\bibitem[Dalcanton \& Hogan(2001)]{2001ApJ...561...35D} Dalcanton, J.~J., \& Hogan, C.~J.\ 2001, ApJ, 561, 35

\bibitem[Dav{\'e}(2008)]{2008MNRAS.385..147D} Dav{\'e}, R.\ 2008, MNRAS, 385, 147

\bibitem[Debattista et al.(2008)]{2008ApJ...681.1076D} Debattista, V.~P., Moore, B., Quinn, T., Kazantzidis, S., Maas, R., Mayer, L., Read, J., \& Stadel, J.\ 2008, ApJ, 681, 1076

\bibitem[Dekel \& Cox(2006)]{2006MNRAS.370.1445D} Dekel, A., \& Cox, T.~J.\ 2006, MNRAS, 370, 1445

\bibitem[Dekel et al.(2009a)]{2009Natur.457..451D} Dekel, A., et al.\ 2009a, Nature, 457, 451

\bibitem[Dekel et al.(2009b)]{2009ApJ...703..785D} Dekel, A., Sari, R., \& Ceverino, D.\ 2009b, ApJ, 703, 785
\bibitem[de Lorenzi et al.(2009)]{2009MNRAS.395...76D} de Lorenzi, F., et al.\ 2009, MNRAS, 395, 76

\bibitem[Donato et al.(2009)]{2009MNRAS.397.1169D} Donato, F., et al.\ 2009, MNRAS, 397, 1169

\bibitem[Duffy et al.(2008)]{2008MNRAS.390L..64D} Duffy, A.~R., Schaye, J., Kay, S.~T., \& Dalla Vecchia, C.\ 2008, MNRAS, 390, L64

\bibitem[Duffy et al.(2010)]{2010arXiv1001.3447D} Duffy, A.~R., Schaye, J., Kay, S.~T., Dalla Vecchia, C., Battye, R.~A., \& Booth, C.~M.\ 2010, MNRAS, submitted, arXiv:1001.3447

\bibitem[Dutton et al.(2007)]{2007ApJ...654...27D} Dutton, A.~A., van den Bosch, F.~C., Dekel, A., \& Courteau, S.\ 2007, ApJ, 654, 27

\bibitem[Elmegreen et al.(2008)]{2008ApJ...688...67E} Elmegreen, B.~G., Bournaud, F., \& Elmegreen, D.~M.\ 2008, ApJ, 688, 67

\bibitem[Ferreras et al.(2005)]{2005ApJ...623L...5F} Ferreras, I., Saha, P., \& Williams, L.~L.~R.\ 2005, ApJ, 623, L5

\bibitem[Ferreras et al.(2008)]{2008MNRAS.383..857F} Ferreras, I., Saha, P., \& Burles, S.\ 2008, MNRAS, 383, 857

\bibitem[Ferreras et al.(2009)]{2009PhRvD..80j3506F} Ferreras, I., Mavromatos, N.~E., Sakellariadou, M., \& Yusaf, M.~F.\ 2009, Phys Rev D, 80, 103506

\bibitem[Forestell(2009)]{2009PhDT.........7F} Forestell, A.~D.\ 2009, Ph.D.~Thesis, Univ. Texas

\bibitem[Forte et al.(2009)]{2009MNRAS.397.1003F} Forte, J.~C., Vega, E.~I., \& Faifer, F.\ 2009, MNRAS, 397, 1003

\bibitem[Foster et al.(2009)]{2009MNRAS.400.2135F} Foster, C., Proctor, R.~N., Forbes, D.~A., Spolaor, M., Hopkins, P.~F., \& Brodie, J.~P.\ 2009, MNRAS, 400, 2135

\bibitem[Frigerio Martins \& Salucci(2007)]{2007MNRAS.381.1103F} Frigerio Martins, C., \& Salucci, P.\ 2007, MNRAS, 381, 1103

\bibitem[Gargiulo et al.(2009)]{2009MNRAS.397...75G} Gargiulo, A., et al.\ 2009, MNRAS, 397, 75

\bibitem[Gavazzi et al.(2007)]{2007ApJ...667..176G} Gavazzi, R., Treu, T., Rhodes, J.~D., Koopmans, L.~V.~E., Bolton, A.~S., Burles, S., Massey, R.~J., \& Moustakas, L.~A.\ 2007, ApJ, 667, 176

\bibitem[Genel et al.(2008)]{2008ApJ...688..789G} Genel, S., et al.\ 2008, ApJ, 688, 789

\bibitem[Gentile et al.(2009)]{2009Natur.461..627G} Gentile, G., Famaey, B., Zhao, H., \& Salucci, P.\ 2009, Nature, 461, 627

\bibitem[Gerhard et al.(2001)]{2001AJ....121.1936G} Gerhard, O., Kronawitter, A., Saglia, R.~P., \& Bender, R.\ 2001, AJ, 121, 1936 (G+01)

\bibitem[Glazebrook et al.(2004)]{2004Natur.430..181G} Glazebrook, K., et al.\ 2004, Nature, 430, 181

\bibitem[\protect\citeauthoryear{Gnedin et al.}{2004}]{Gnedin+04} Gnedin, O.~Y., Kravtsov,
A.~V., Klypin, A.~A., \& Nagai, D.\ 2004, ApJ, 616, 16 (G+04)

\bibitem[Gnedin et al.(2007)]{2007ApJ...671.1115G} Gnedin, O.~Y., Weinberg, D.~H., Pizagno, J., Prada, F., \& Rix, H.-W.\ 2007, ApJ, 671, 1115

\bibitem[Governato et al.(2010)]{2010Natur.463..203G} Governato, F., et al.\ 2010, Nature, 463, 203

\bibitem[Graham et al.(2006a)]{2006AJ....132.2701G} Graham, A.~W., Merritt, D., Moore, B., Diemand, J., \& Terzi{\'c}, B.\ 2006a, AJ, 132, 2701

\bibitem[Graham et al.(2006b)]{2006AJ....132.2711G} Graham, A.~W., Merritt, D., Moore, B., Diemand, J., \& Terzi{\'c}, B.\ 2006b, AJ, 132, 2711

\bibitem[Graves et al.(2009a)]{2009ApJ...693..486G} Graves, G.~J., Faber, S.~M., \& Schiavon, R.~P.\ 2009a, ApJ, 693, 486

\bibitem[Graves et al.(2009b)]{2009ApJ...698.1590G} Graves, G.~J., Faber, S.~M., \& Schiavon, R.~P.\ 2009b, ApJ, 698, 1590

\bibitem[Graves(2009)]{2009PhDT.........9G} Graves, G.~J.\ 2009, Ph.D.~Thesis, Univ. California, Santa Cruz

\bibitem[Grillo \& Gobat(2010)]{2010MNRAS.402L..67G} Grillo, C., \& Gobat, R.\ 2010, MNRAS, 402, L67

\bibitem[Guo et al.(2010)]{2009arXiv0909.4305G} Guo, Q., White, S., Li, C., \& Boylan-Kolchin, M.\ 2010, MNRAS, submitted, arXiv:0909.4305

\bibitem[Gustafsson et al.(2006)]{2006PhRvD..74l3522G} Gustafsson, M., Fairbairn, M., \& Sommer-Larsen, J.\ 2006, Phys Rev D, 74, 123522

\bibitem[Haas \& Anders(2010)]{2010arXiv1001.2009H} Haas, M.~R., \& Anders, P.\ 2010, A\&A, submitted, arXiv:1001.2009

\bibitem[Hernquist et al.(1993)]{1993ApJ...416..415H} Hernquist, L., Spergel, D.~N., \& Heyl, J.~S.\ 1993, ApJ, 416, 415

\bibitem[Hoffman et al.(2010)]{2010arXiv1001.0799H} Hoffman, L., Cox, T.~J., Dutta, S., \& Hernquist, L.\ 2010, ApJ, submitted, arXiv:1001.0799

\bibitem[Holden et al.(2010)]{Holden10} Holden, B., et al. 2010, ApJ, submitted

\bibitem[Hopkins et al.(2008)]{2008ApJ...689...17H} Hopkins, P.~F., Cox, T.~J., \& Hernquist, L.\ 2008, ApJ, 689, 17

\bibitem[Hopkins et al.(2010)]{2010MNRAS.401.1099H} Hopkins, P.~F., Bundy, K., Hernquist, L., Wuyts, S., \& Cox, T.~J.\ 2010, MNRAS, 401, 1099

\bibitem[Hopkins \& Hernquist(2010)]{2010MNRAS.402..985H} Hopkins, P.~F., \& Hernquist, L.\ 2010, MNRAS, 402, 985

\bibitem[Humphrey et al.(2009)]{2009ApJ...703.1257H} Humphrey, P.~J., Buote, D.~A., Brighenti, F., Gebhardt, K., \& Mathews, W.~G.\ 2009, ApJ, 703, 1257

\bibitem[Humphrey \& Buote(2010)]{2010MNRAS.tmp..135H} Humphrey, P.~J., \& Buote, D.~A.\ 2010, MNRAS, 135

\bibitem[Jardel \& Sellwood(2009)]{2009ApJ...691.1300J} Jardel, J.~R., \& Sellwood, J.~A.\ 2009, ApJ, 691, 1300

\bibitem[Jesseit et al.(2009)]{2009MNRAS.397.1202J} Jesseit, R., Cappellari, M., Naab, T., Emsellem, E., \& Burkert, A.\ 2009, MNRAS, 397, 1202

\bibitem[Jiang \& Kochanek(2007)]{2007ApJ...671.1568J} Jiang, G., \& Kochanek, C.~S.\ 2007, ApJ, 671, 1568

\bibitem[Johansson et al.(2009)]{2009ApJ...697L..38J} Johansson, P.~H., Naab, T., \& Ostriker, J.~P.\ 2009, ApJ, 697, L38

\bibitem[Jun \& Im(2008)]{2008ApJ...678L..97J} Jun, H.~D., \& Im, M.\ 2008, ApJ, 678, L97

\bibitem[Kalirai et al.(2010)]{2009arXiv0911.1998K} Kalirai, J.~S., et al.\ 2010, ApJ, in pres, arXiv:0911.1998

\bibitem[Kassin et al.(2006)]{2006ApJ...643..804K} Kassin, S.~A., de Jong, R.~S., \& Weiner, B.~J.\ 2006, ApJ, 643, 804

\bibitem[Kauffmann et al.(2003)]{2003MNRAS.346.1055K} Kauffmann, G., et al.\ 2003, MNRAS, 346, 1055

\bibitem[Kazantzidis et al.(2006)]{2006ApJ...641..647K} Kazantzidis, S., Zentner, A.~R., \& Kravtsov, A.~V.\ 2006, ApJ, 641, 647

\bibitem[Kere{\v s} et al.(2009a)]{2009MNRAS.395..160K} Kere{\v s}, D., Katz, N., Fardal, M., Dav{\'e}, R., \& Weinberg, D.~H.\ 2009a, MNRAS, 395, 160

\bibitem[Kere{\v s} et al.(2009b)]{2009MNRAS.396.2332K} Kere{\v s}, D., Katz, N., Dav{\'e}, R., Fardal, M., \& Weinberg, D.~H.\ 2009b, MNRAS, 396, 2332

\bibitem[Klessen et al.(2007)]{2007MNRAS.374L..29K} Klessen, R.~S., Spaans, M., \& Jappsen, A.-K.\ 2007, MNRAS, 374, L29

\bibitem[Koopmans et al.(2009)]{2009ApJ...703L..51K} Koopmans, L.~V.~E., et al.\ 2009, ApJ, 703, L51

\bibitem[Koposov et al.(2009)]{2009ApJ...696.2179K} Koposov, S.~E., Yoo, J., Rix, H.-W., Weinberg, D.~H., Macci{\`o}, A.~V., \& Escud{\'e}, J.~M.\ 2009, ApJ, 696, 2179

\bibitem[Kormendy \& Freeman(2004)]{2004IAUS..220..377K} Kormendy, J., \& Freeman, K.~C.\ 2004, Dark Matter in Galaxies, 220, 377

\bibitem[Kormendy \& Sanders(1992)]{1992ApJ...390L..53K} Kormendy, J., \& Sanders, D.~B.\ 1992, ApJ, 390, L53

\bibitem[\protect\citeauthoryear{Kroupa}{2001}]{Kroupa01}Kroupa P., 2001, MNRAS, 322, 231

\bibitem[Krumholz et al.(2010)]{2010arXiv1001.0971K} Krumholz, M.~R., Cunningham, A.~J., Klein, R.~I., \& McKee, C.~F.\ 2010, ApJ, submitted, arXiv:1001.0971

\bibitem[La Barbera et al.(2010a)]{2009arXiv0912.4558L} La Barbera, F., de Carvalho, R.~R., de la Rosa, I.~G., \& Lopes, P.~A.~A.\ 2010a, MNRAS, submitted, arXiv:0912.4558

\bibitem[La Barbera et al.(2010b)]{2010arXiv1003.1119L} La Barbera, F., Lopes, P.~A.~A., de Carvalho, R.~R., de la Rosa, I.~G., \& Berlind, A.~A.\ 2010b, MNRAS, submitted, arXiv:1003.1119

\bibitem[Lackner \& Ostriker(2010)]{2010ApJ...712...88L} Lackner, C.~N., \& Ostriker, J.~P.\ 2010, ApJ, 712, 88

\bibitem[Lagattuta et al.(2010)]{2009arXiv0911.2236L} Lagattuta, D.~J., et al.\ 2010, ApJ, submitted, arXiv:0911.2236

\bibitem[Larson(2005)]{2005MNRAS.359..211L} Larson, R.~B.\ 2005, MNRAS, 359, 211

\bibitem[Lee et al.(2004)]{2004MNRAS.353..113L} Lee, H.-c., Gibson, B.~K., Flynn, C., Kawata, D., \& Beasley, M.~A.\ 2004, MNRAS, 353, 113

\bibitem[Li et al.(2009)]{2009MNRAS.397L..87L} Li, Y.-S., Helmi, A., De Lucia, G., \& Stoehr, F.\ 2009, MNRAS, 397, L87

\bibitem[Macci{\`o} et al.(2008)]{2008MNRAS.391.1940M} Macci{\`o}, A.~V., Dutton, A.~A., \& van den Bosch, F.~C.\ 2008, MNRAS, 391, 1940

\bibitem[Macci{\`o} et al.(2009)]{2009ApJ...692L.109M} Macci{\`o}, A.~V., Kang, X., \& Moore, B.\ 2009, ApJ, 692, L109

\bibitem[Mancini et al.(2010)]{2010MNRAS.401..933M} Mancini, C., et al.\ 2010, MNRAS, 401, 933

\bibitem[\protect\citeauthoryear{Mashchenko et al.}{2008}]{Mash+08} Mashchenko, S., Wadsley, J., \& Couchman, H.~M.~P.\ 2008, Science, 319, 174

\bibitem[McGaugh(2005)]{2005ApJ...632..859M} McGaugh, S.~S.\ 2005, ApJ, 632, 859

\bibitem[McGaugh et al.(2007)]{2007ApJ...659..149M} McGaugh, S.~S., de Blok, W.~J.~G., Schombert, J.~M., Kuzio de Naray, R., \& Kim, J.~H.\ 2007, ApJ, 659, 149

\bibitem[McGaugh(2008)]{2008ApJ...683..137M} McGaugh, S.~S.\ 2008, ApJ, 683, 137

\bibitem[Meurer et al.(2009)]{2009ApJ...695..765M} Meurer, G.~R., et al.\ 2009, ApJ, 695, 765

\bibitem[Mao \& Mo(1998)]{1998MNRAS.296..847M} Mao, S., \& Mo, H.~J.\ 1998, MNRAS, 296, 847

\bibitem[Mo \& Mao(2004)]{2004MNRAS.353..829M} Mo, H.~J., \& Mao, S.\ 2004, MNRAS, 353, 829

\bibitem[More et al.(2009)]{2009MNRAS.392..801M} More, S., van den Bosch, F.~C., Cacciato, M., Mo, H.~J., \& Yang, X., \& Li, R.\ 2009, MNRAS, 392, 801

\bibitem[More et al.(2010)]{2010More} More, S., van den Bosch, F.~C., Cacciato, M., Skibba, R., Mo, H.~J., \& Yang, X.\ 2010, MNRAS, in prep.

\bibitem[Moster et al.(2010)]{2010ApJ...710..903M} Moster, B.~P., Somerville, R.~S., Maulbetsch, C., van den Bosch, F.~C., Macci{\`o}, A.~V., Naab, T., \& Oser, L.\ 2010, ApJ, 710, 90

\bibitem[Muzzin et al.(2009)]{2009ApJ...706L.188M} Muzzin, A., van Dokkum, P., Franx, M., Marchesini, D., Kriek, M., \& Labb{\'e}, I.\ 2009, ApJ, 706, L188

\bibitem[Naab et al.(2009)]{2009ApJ...699L.178N} Naab, T., Johansson, P.~H., \& Ostriker, J.~P.\ 2009, ApJ, 699, L178

\bibitem[\protect\citeauthoryear{Napolitano et al.}{2005}]{Nap05} Napolitano, N. R., et al.
2005, MNRAS, 357, 691

\bibitem[\protect\citeauthoryear{Napolitano et al.}{2009}]{Nap09} Napolitano, N. R.,
et al. 2009, MNRAS, 393, 329 (N+09)

\bibitem[Navarro et al.(1997)]{1997ApJ...490..493N} Navarro, J.~F., Frenk, C.~S., \& White, S.~D.~M.\ 1997, ApJ, 490, 493

\bibitem[Niemi et al.(2010)]{2010arXiv1002.0847N} Niemi, S.-M., Hein{\"a}m{\"a}ki, P., Nurmi, P., \& Saar, E.\ 2010, MNRAS, in press, arXiv:1002.0847

\bibitem[Nipoti et al.(2009)]{2009ApJ...703.1531N} Nipoti, C., Treu, T., \& Bolton, A.~S.\ 2009, ApJ, 703, 1531

\bibitem[Okamoto \& Frenk(2009)]{2009MNRAS.399L.174O} Okamoto, T., \& Frenk, C.~S.\ 2009, MNRAS, 399, L174

\bibitem[O{\~n}orbe et al.(2005)]{2005ApJ...632L..57O} O{\~n}orbe, J., Dom{\'{\i}}nguez-Tenreiro, R., S{\'a}iz, A., Serna, A., \& Artal, H.\ 2005, ApJ, 632, L57

\bibitem[O{\~n}orbe et al.(2006)]{2006MNRAS.373..503O} O{\~n}orbe, J., Dom{\'{\i}}nguez-Tenreiro, R., S{\'a}iz, A., Artal, H., \& Serna, A.\ 2006, MNRAS, 373, 503

\bibitem[Pedrosa et al.(2009)]{2009MNRAS.395L..57P} Pedrosa, S., Tissera, P.~B., \& Scannapieco, C.\ 2009, MNRAS, 395, L57

\bibitem[Pedrosa et al.(2010)]{2010MNRAS.402..776P} Pedrosa, S., Tissera, P.~B., \& Scannapieco, C.\ 2010, MNRAS, 402, 776

\bibitem[Peirani et al.(2008)]{2008A&A...479..123P} Peirani, S., Kay, S., \& Silk, J.\ 2008, A\&A, 479, 123

\bibitem[Pe\~narrubia et al.(2010)]{2010arXiv1002.3376P} Pe\~narrubia, J., Benson, A.~J., Walker, M.~G., Gilmore, G., McConnachie, A., \& Mayer, L.\ 2010, MNRAS, submitted, arXiv:1002.3376

\bibitem[Pointecouteau \& Silk(2005)]{2005MNRAS.364..654P} Pointecouteau, E., \& Silk, J.\ 2005, MNRAS, 364, 654

\bibitem[Proctor et al.(2009)]{2009MNRAS.398...91P} Proctor, R.~N., Forbes, D.~A., Romanowsky, A.~J., Brodie, J.~P., Strader, J., Spolaor, M., Mendel, J.~T., \& Spitler, L.\ 2009, MNRAS, 398, 91

\bibitem[\protect\citeauthoryear{Prugniel \& Simien}{1996}]{PS96} Prugniel, Ph. \& Simien F. 1996, A\&A, 309, 749

\bibitem[Renzini(2005)]{2005ASSL..327..221R} Renzini, A.\ 2005, The Initial Mass Function 50 Years Later, 327, 221

\bibitem[Renzini(2006)]{2006ARA&A..44..141R} Renzini, A.\ 2006, ARAA, 44, 141

\bibitem[Rhode et al.(2007)]{2007AJ....134.1403R} Rhode, K.~L., Zepf, S.~E., Kundu, A., \& Larner, A.~N.\ 2007, AJ, 134, 1403

\bibitem[Richtler et al.(2008)]{2008A&A...478L..23R} Richtler, T., Schuberth, Y., Hilker, M., Dirsch, B., Bassino, L., \& Romanowsky, A.~J.\ 2008, A\&A, 478, L23

\bibitem[Robertson et al.(2006)]{2006ApJ...641...21R} Robertson, B., Cox, T.~J., Hernquist, L., Franx, M., Hopkins, P.~F., Martini, P., \& Springel, V.\ 2006, ApJ, 641, 21

\bibitem[Rogers et al.(2010)]{2010arXiv1002.0835R} Rogers, B., Ferreras, I., Pasquali, A., Bernardi, M., Lahav, O., \& Kaviraj, S.\ 2010, MNRAS, in press, arXiv:1002.0835

\bibitem[Romanowsky(2006)]{2006EAS....20..119R} Romanowsky, A.~J.\ 2006, EAS Publications Series, 20, 119

\bibitem[Romanowsky et al.(2009)]{2009AJ....137.4956R} Romanowsky, A.~J., Strader, J., Spitler, L.~R., Johnson, R., Brodie, J.~P., Forbes, D.~A., \& Ponman, T.\ 2009, AJ, 137, 4956

\bibitem[Romano-D{\'{\i}}az et al.(2008)]{2008ApJ...685L.105R} Romano-D{\'{\i}}az, E., Shlosman, I., Hoffman, Y., \& Heller, C.\ 2008, ApJ, 685, L105

\bibitem[Ruszkowski \& Springel(2009)]{2009ApJ...696.1094R} Ruszkowski, M., \& Springel, V.\ 2009, ApJ, 696, 1094

\bibitem[\protect\citeauthoryear{Salpeter}{1955}]{Salpeter55} Salpeter, E.E. 1955 ApJ, 121, 161

\bibitem[S{\'a}nchez-Bl{\'a}zquez et al.(2007)]{2007MNRAS.377..759S} S{\'a}nchez-Bl{\'a}zquez, P., Forbes, D.~A., Strader, J., Brodie, J., \& Proctor, R.\ 2007, MNRAS, 377, 759

\bibitem[Sanders(2003)]{2003MNRAS.342..901S} Sanders, R.~H.\ 2003, MNRAS, 342, 901

\bibitem[Sanders \& Land(2008)]{2008MNRAS.389..701S} Sanders, R.~H., \& Land, D.~D.\ 2008, MNRAS, 389, 701

\bibitem[Saracco et al.(2009)]{2009MNRAS.392..718S} Saracco, P., Longhetti, M., \& Andreon, S.\ 2009, MNRAS, 392, 718

\bibitem[Saxton \& Ferreras(2010)]{2010arXiv1002.0845S} Saxton, C.~J., \& Ferreras, I.\ 2010, MNRAS, in press, arXiv:1002.0845

\bibitem[Schmitt et al.(1999)]{1999MNRAS.303..173S} Schmitt, H.~R., Storchi-Bergmann, T., \& Cid Fernandes, R.\ 1999, MNRAS, 303, 173

\bibitem[Schulz et al.(2010)]{2009arXiv0911.2260S} Schulz, A.~E., Mandelbaum, R., \& Padmanabhan, N.\ 2010, MNRAS, submitted, arXiv:0911.2260 (S+10)

\bibitem[Sellwood \& McGaugh(2005)]{2005ApJ...634...70S} Sellwood, J.~A., \& McGaugh, S.~S.\ 2005, ApJ, 634, 70

\bibitem[\protect\citeauthoryear{S\'ersic}{1968}]{Sersic68} S\'ersic, J. L. 1968, Atlas de Galaxies Australes, Observatorio Astronomico de Cordoba

\bibitem[Shankar \& Bernardi(2009)]{2009MNRAS.396L..76S} Shankar, F., \& Bernardi, M.\ 2009, MNRAS, 396, L76 (SB09)

\bibitem[Shankar et al.(2010)]{2009arXiv0912.0012S} Shankar, F., Marulli, F., Bernardi, M., Dai, X., Hyde, J.~B., \& Sheth, R.~K.\ 2010, MNRAS, in press, arXiv:0912.0012

\bibitem[Shen \& Gebhardt(2010)]{2009arXiv0910.4168S} Shen, J., \& Gebhardt, K.\ 2010, ApJ, in press, arXiv:0910.4168

\bibitem[Shin \& Kawata(2009)]{2009ApJ...691...83S} Shin, M.-S., \& Kawata, D.\ 2009, ApJ, 691, 83

\bibitem[Sommer-Larsen \& Toft(2010)]{2009arXiv0909.0943S} Sommer-Larsen, J., \& Toft, S.\ 2010, ApJ, submitted, arXiv:0909.0943

\bibitem[\protect\citeauthoryear{Spergel et al.}{2007}]{WMAP} Spergel, D.~N., et al.,  2007, ApJS, 170, 377

\bibitem[Spolaor et al.(2008)]{2008MNRAS.385..675S} Spolaor, M., Forbes, D.~A., Proctor, R.~N., Hau, G.~K.~T., \& Brough, S.\ 2008, MNRAS, 385, 675

\bibitem[Springel(2000)]{2000MNRAS.312..859S} Springel, V.\ 2000, MNRAS, 312, 859

\bibitem[Strigari et al.(2008)]{2008Natur.454.1096S} Strigari, L.~E., Bullock, J.~S., Kaplinghat, M., Simon, J.~D., Geha, M., Willman, B., \& Walker, M.~G.\ 2008, Nature, 454, 1096

\bibitem[Stringer et al.(2010)]{2009arXiv0911.1888S} Stringer, M., Cole, S., \& Frenk, C.\ 2010, arXiv:0911.1888

\bibitem[Terlevich \& Forbes(2002)]{2002MNRAS.330..547T} Terlevich, A.~I., \& Forbes, D.~A.\ 2002, MNRAS, 330, 547

\bibitem[\protect\citeauthoryear{Thomas et al.}{2005}]{Thomas+05} Thomas, D., Maraston, C., Bender, R. \& Mendes de Oliveira, C. 2005, ApJ, 621, 673

\bibitem[Thomas et al.(2007)]{2007MNRAS.382..657T} Thomas, J., Saglia, R.~P., Bender, R., Thomas, D., Gebhardt, K., Magorrian, J., Corsini, E.~M., \& Wegner, G.\ 2007, MNRAS, 382, 657

\bibitem[Thomas et al.(2009)]{2009ApJ...691..770T} Thomas, J., Saglia, R.~P., Bender, R., Thomas, D., Gebhardt, K., Magorrian, J., Corsini, E.~M., \& Wegner, G.\ 2009, ApJ, 691, 770 (T+09)

\bibitem[Tiret et al.(2007)]{2007A&A...476L...1T} Tiret, O., Combes, F., Angus, G.~W., Famaey, B., \& Zhao, H.~S.\ 2007, A\&A, 476, L1

\bibitem[Tissera et al.(2010)]{2009arXiv0911.2316T} Tissera, P.~B., White, S.~D.~M., Pedrosa, S., \& Scannapieco, C.\ 2010, MNRAS, submitted, arXiv:0911.2316

\bibitem[Tortora et al.(2009)]{2009MNRAS.396.1132T} Tortora, C., Napolitano, N.~R., Romanowsky, A.~J., Capaccioli, M., \& Covone, G.\ 2009, MNRAS, 396, 1132 (paper I)

\bibitem[Treu et al.(2010)]{2010ApJ...709.1195T} Treu, T., Auger, M.~W., Koopmans, L.~V.~E., Gavazzi, R., Marshall, P.~J., \& Bolton, A.~S.\ 2010, ApJ, 709, 1195 (T+10)

\bibitem[Trujillo et al.(2004)]{2004ApJ...600L..39T} Trujillo, I., Burkert, A., \& Bell, E.~F.\ 2004, ApJ, 600, L39

\bibitem[Trujillo et al.(2006)]{2006ApJ...650...18T} Trujillo, I., et al.\ 2006, ApJ, 650, 18

\bibitem[Valentinuzzi et al.(2010)]{2009arXiv0907.2392V} Valentinuzzi, T., et al.\ 2010, ApJ, in press, arXiv:0907.2392


\bibitem[van der Wel et al.(2009)]{2009ApJ...698.1232V} van der Wel, A., Bell, E.~F., van den Bosch, F.~C., Gallazzi, A., \& Rix, H.-W.\ 2009, ApJ, 698, 1232

\bibitem[van Dokkum(2008)]{2008ApJ...674...29V} van Dokkum, P.~G.\ 2008, ApJ, 674, 29

\bibitem[van Dokkum et al.(2009)]{2009Natur.460..717V} van Dokkum, P.~G., Kriek, M., \& Franx, M.\ 2009, Nature, 460, 717

\bibitem[van Dokkum et al.(2010)]{2010ApJ...709.1018V} van Dokkum, P.~G., et al.\ 2010, ApJ, 709, 1018

\bibitem[Walker et al.(2009)]{2009ApJ...704.1274W} Walker, M.~G., Mateo, M., Olszewski, E.~W., Pe{\~n}arrubia, J., Wyn Evans, N., \& Gilmore, G.\ 2009, ApJ, 704, 1274

\bibitem[Walker et al.(2010)]{2010ApJ...710..886W} Walker, M.~G., Mateo, M., Olszewski, E.~W., Pe{\~n}arrubia, J., Wyn Evans, N., \& Gilmore, G.\ 2010, ApJ, 710, 886

\bibitem[Wechsler et al.(2002)]{2002ApJ...568...52W} Wechsler, R.~H., Bullock, J.~S., Primack, J.~R., Kravtsov, A.~V., \& Dekel, A.\ 2002, ApJ, 568, 52

\bibitem[Wechsler et al.(2006)]{2006ApJ...652...71W} Wechsler, R.~H., Zentner, A.~R., Bullock, J.~S., Kravtsov, A.~V., \& Allgood, B.\ 2006, ApJ, 652, 71

\bibitem[Weidner \& Kroupa(2006)]{2006MNRAS.365.1333W} Weidner, C., \& Kroupa, P.\ 2006, MNRAS, 365, 1333

\bibitem[Weijmans et al.(2009)]{2009MNRAS.398..561W} Weijmans, A.-M., et al.\ 2009, MNRAS, 398, 561

\bibitem[Wolf et al.(2010)]{2009arXiv0908.2995W} Wolf, J., Martinez, G.~D., Bullock, J.~S., Kaplinghat, M., Geha, M., Munoz, R.~R., Simon, J.~D., \& Avedo, F.~F.\ 2010, MNRAS, submitted arXiv:0908.2995

\bibitem[Woodley et al.(2010)]{2010arXiv1002.3142W} Woodley, K.~A., Gomez, M., Harris, W.~E., Geisler, D., \& Harris, G.~L.~H.\ 2010, AJ, in press, arXiv:1002.3142

\bibitem[Xue et al.(2008)]{2008ApJ...684.1143X} Xue, X.~X., et al.\ 2008, ApJ, 684, 1143

\bibitem[Zhao et al.(2008)]{2008ApJ...686.1019Z} Zhao, H., Xu, B.-X., \& Dobbs, C.\ 2008, ApJ, 686, 1019

\bibitem[Zheng et al.(2007)]{2007ApJ...667..760Z} Zheng, Z., Coil, A.~L., \& Zehavi, I.\ 2007, ApJ, 667, 760

\end{thebibliography}
\end{document}